\documentclass[aps, pra, reprint, superscriptaddress, dvipdfmx]{revtex4-1}

\usepackage{amsmath, amssymb, bm}
\usepackage{graphicx}

\usepackage{color}
\usepackage[colorlinks=true, linkcolor=magenta, citecolor=blue, urlcolor=magenta]{hyperref}

\newtheorem{conjecture}{Conjecture}

\newtheorem{corollaryConjecture}{Corollary of Conjecture}

\begin{document}

\title{Superposition of Macroscopically Distinct States in Adiabatic Quantum Computation}

\author{Tatsuro Yuge}
\affiliation{Department of Physics, Shizuoka University, Suruga, Shizuoka 422-8529, Japan}

\date{\today}

\begin{abstract}
What are the conditions for adiabatic quantum computation (AQC) to outperform classical computation?
Although there exist several quantum adiabatic algorithms achieving the strong quantum speedup,
the essential keys to their speedups are still unclear.
Here, we investigate the connection between superpositions of macroscopically distinct states
and known examples of the speedup in AQC.
To formalize this notion we consider an index $p$ that quantifies a superposition of macroscopically distinct states from the asymptotic behaviors of fluctuations of additive observables.
We determine this index for five examples of adiabatic algorithms exhibiting various degrees
of the speedup.
The results suggest that the superposition of macroscopically distinct states
is an appropriate indicator of entanglement crucial to the strong quantum speedup in AQC.
\end{abstract}


\maketitle

\section{Introduction}

Adiabatic quantum computation (AQC) can outperform classical computation.
But why and in what conditions does it achieve the quantum speedup?
This fundamental question is yet to be answered although,
since the proposal of AQC \cite{Farhi_etal:2000, Farhi_etal:2001} and quantum annealing
\cite{Apolloni_etal:1989, Somorjai:1991, Amara_etal:1993, Finnila_etal:1994, Kadowaki_etal:1998},
intensive studies have been carried out
\cite{Santoro_etal:2006, Das_etal:2008, Morita_etal:2008, Bapst_etal:2013, Suzuki_etal:2015, Albash_etal:2016}.

The basic procedure of AQC for optimization problems is as follows.
First, we encode the cost function of a problem instance into a ``problem Hamiltonian'' $\hat{H}_p(n,\nu)$
such that its ground state minimizes the cost function.
Here, $n$ is the problem size and $\nu$ is a symbolic label of instances.
Second, we prepare a system (a quantum device) in the ground state of another Hamiltonian $\hat{H}_d(n)$,
called ``driver Hamiltonian.''
$\hat{H}_d$ should not commute with $\hat{H}_p$, and $\hat{H}_d$'s ground state should be easily prepared.
Third, we let the system evolve according to the Shr\"odinger equation
with the following time-dependent Hamiltonian:
\begin{align}
\hat{H}(s;n,\nu) = s \hat{H}_p(n,\nu) + (1-s) \hat{H}_d(n).
\label{Hs}
\end{align}
Here, we vary the ``annealing parameter'' $s = s(t)$ in time $t$ from $s(0) = 0$ to $s(T) = 1$.
If the rate of change $\dot{s}(t)$ is sufficiently small to satisfy the adiabatic condition
\cite{Kato:1950, Messiah} (see also Sec.~II of Ref.~\cite{Albash_etal:2016}),
then the state stays close to the instantaneous ground state of $\hat{H}(s;n,\nu)$ at any time.
And, at the end ($t = T$) of the time evolution,
we obtain the ground state of $\hat{H}_p(n,\nu)$, the solution of the instance.
From the adiabatic theorem, the computational time $T$
depends on a polynomial of the inverse of the energy gap between the ground and first excited states of $\hat{H}(s;n,\nu)$.

There are quantum adiabatic algorithms which do outperform classical computation
in the sense of the strong quantum speedup \cite{Papageorgiou_etal:2013, Ronnow_etal:2014}
(the scaling advantage compared with the best classical algorithms).
One of them is the adiabatic algorithm for Grover's search problem \cite{Roland_etal:2002}.
It was shown that its computational time is the same order
as that of the original Grover algorithm \cite{Grover:1997}.
And several other algorithms were found \cite{Somma_etal:2012, Hen:2014a, Hen:2014b, Hen:2016, Albash_etal:2016}.
However, in general and in many problems, it is unclear whether AQC achieves the quantum speedup or not.
In addition, even if restricted to the algorithms known to outperform,
it is unclear what features of quantum physics are essential to their quantum speedups.
For example, although quantum tunneling is often expected to be a key to the speedup of AQC,
there are counterexamples to this expectation \cite{Muthukrishnan_etal:2016}.

Clarifying the mechanism and conditions of quantum speedup is crucial for developing AQC.
They will enable us to judge whether a given algorithm of AQC achieves quantum speedup.
They will be also useful for real quantum devices,
such as D-wave machines \cite{Johnson_etal:2011, Lanting_etal:2014, Ronnow_etal:2014, Boixo_etal:2014, Katzgraber:2015, Denchev_etal:2016},
which implement algorithms of quantum annealing and AQC:
we can judge quantum speedup of algorithms on devices
by investigating whether they satisfy the conditions,
instead of by direct comparison with classical ones.

As a quantum effect possibly connected to the quantum speedup,
the many-body localization may be considered.
This phenomenon induces an exponentially small energy gap
and is thus considered as a bottleneck for AQC \cite{Altshuler_etal:2010, Bapst_etal:2013}.
And methods of circumventing this phenomenon have been argued
\cite{Farhi_etal:2011, Choi:2010, Dickson_etal:2011, Dickson:2011}.
But its relation to the strong quantum speedup is not clear
because some algorithms achieve the speedup even with exponentially small gaps.

As another quantum effect, we here consider quantum entanglement.
In circuit-based quantum computation, many studies assert the necessity of entanglement for the quantum speedup
\cite{Jozsa_etal:2003, Vidal:2003, Parker_etal:2002, Shimoni_etal:2004, Shimoni_etal:2005, Orus_etal:2004, Ukena_etal:2004, Shimizu_etal:2013}.
In AQC, by contrast, the necessity of entanglement is less clear although there are several theoretical studies
\cite{Orus_etal:2004, Latorre_etal:2004, Rezakhani_etal:2009, Bauer_etal:2015, Hauke_etal:2015, Batle_etal:2016}
and although entanglement is experimentally detected in the D-wave machine
\cite{Lanting_etal:2014} (see also Ref.~\cite{Albash_etal:2015}).

Or\'us and Latorre studied the entanglement entropy in AQC \cite{Orus_etal:2004, Latorre_etal:2004}:
in the exact cover problem
a simulation with up to 20 qubits showed that
the entanglement entropy increases proportionally to the system size,
whereas in the adiabatic Grover algorithm an analytical calculation showed
that the entanglement entropy is bounded (approaches a constant value as the size increases).
Rezakhani \textit{et al.} showed that
entanglement (negativity) in the optimal AQC algorithm for Grover's search problem
is smaller than that in a non-optimal case \cite{Rezakhani_etal:2009}.
Relation between entanglement and the success probability of finding the solution was argued in simulations of AQC \cite{Bauer_etal:2015, Hauke_etal:2015}.
In Ref.~\cite{Bauer_etal:2015}, it was reported that,
in an AQC  algorithm for Ising spin glass problems,
the success probability increases as the limit of entanglement allowed in the simulation is increased.
In Ref.~\cite{Hauke_etal:2015}, by contrast, it was reported that, in an AQC algorithm for the Coulomb-glass problem,
the entanglement entropy of the intermediate states has little significance for the success probability
although that of the final state provides an upper bound on the probability.
Reference \cite{Batle_etal:2016} investigated multipartite entanglement
in AQC algorithms for the factoring problem and the Deutsch--Jozsa problem.
It reported that no significantly large entanglement appears during computation.

In the present paper, we propose a necessary condition for the strong quantum speedup in AQC.
We conjecture that a superposition of macroscopically distinct states appears during AQC,
if it achieves the speedup.
This conjecture is an extension of that proposed in a context of circuit-based quantum computation
\cite{Ukena_etal:2004, Shimizu_etal:2013}.
We investigate whether our conjecture is correct in five AQC algorithms:
four of them are known to achieve the quantum speedup and the rest one is known not to achieve.
As summarized in Table~\ref{table:results}, all of the results support our conjecture.

\begin{table}[bt]
\caption{
Results of the tests for our conjecture (see Sec.~\ref{sec:examples} for details).
The cited articles are those which proposed the AQC algorithms investigated here.
In the middle column, described are known results on the strong quantum speedup.
In the right column, $p = 2$ means that a superposition of macroscopically distinct states
appears somewhere in the computation
while $p = 1$ means that none of the states during the computation are
superpositions of macroscopically distinct states.
See Sec.~\ref{sec:index_p} for the definition of the index $p$,
an indicator of a superposition of macroscopically distinct states.
}
\label{table:results}
\begin{ruledtabular}
\begin{tabular}{lcl}
& Strong quantum speedup & $p$
\\
\colrule
Grover \cite{Roland_etal:2002} & Achieved & 2
\\
Deutsch--Jozsa \cite{Wei_etal:2006} & Not achieved & 1
\\
Bernstein--Vazirani \cite{Hen:2014b} & Achieved & 2
\\
Simon \cite{Hen:2014b} & Achieved & 2
\\
Glued-trees \cite{Somma_etal:2012} & Achieved & 2
\\
\end{tabular}
\end{ruledtabular}
\end{table}

The structure of the present paper is as follows.
In Sec.~\ref{sec:definitions}, we review a superposition of macroscopically distinct states.
We mention the conjecture in Sec.~\ref{sec:conjecture}
and show its plausibility by testing it in several algorithms in Sec.~\ref{sec:examples}.
In Sec.~\ref{sec:conclusion}, we make concluding remarks.
We describe some technical details in Appendices.

\section{Superposition of macroscopically distinct states}\label{sec:definitions}

In this section, we briefly review a superposition of macroscopically distinct states and its quantification.
Suppose a quantum system composed of $n$ local sites.
We assume that the dimension of the Hilbert space $\mathcal{H}_\mathrm{loc}$ of each site is finite
and any local operators on $\mathcal{H}_\mathrm{loc}$ are bounded.
Here, we consider only pure states because we discuss AQC that uses only pure states in this study.
For more details, see Ref.~\cite{Shimizu_etal:2013}.

\subsection{Macroscopically distinct states and their superposition}

We first define ``macroscopically distinct states.''
Two or more states are macroscopically distinct if there is at least one macroscopic quantity
whose values are different among them in macroscopic accuracy.
There are many macroscopic quantities---total energy, magnetization, temperature, entropy, and so on.
Among them we take ``mechanical additive observables'' (such as total energy and magnetization).
They are expressed as sums of local operators.
In this choice, the macroscopic accuracy means resolution of $O(n)$.

Therefore, we can say that
two pure states, $|\psi_1 \rangle$ and $|\psi_2 \rangle$, are macroscopically distinct
if $\big| \langle \psi_1 | \hat{A} |\psi_1 \rangle - \langle \psi_2 | \hat{A} |\psi_2 \rangle \big| = O(n)$
for an additive operator $\hat{A}$.
An additive operator $\hat{A}$ is defined as $\hat{A} \equiv \sum_{l=1}^n \hat{a}(l)$,
where $\hat{a}(l)$ is a local operator at the $l$th site and $l$ runs over all the sites in the system.
We assume that $|\hat{a}(l)| = O(1)$ as usual so that $|\hat{A}| = O(n)$,
where the operator norm is defined appropriately.

We next explain a superposition of macroscopically distinct states.
To this end we first consider a superposition of macroscopically \textit{identical} (non-distinct) states.
Let $\hat{A}$ be any additive operator,
and let $|A_i \rangle$ be an eigenstate of $\hat{A}$ with an eigenvalue $A_i$.
From the above argument,
any superposition $|\phi \rangle$ of macroscopically identical states contains almost only $|A_i \rangle$'s
that satisfy $\big| A - A_i \big| = o(n)$ with an $i$-independent value $A$.
In other words, the probability to find $|A_i \rangle$'s with $\big| A - A_i \big| = O(n)$ is vanishingly small.
Then, we can show that $|\phi \rangle$ has normal fluctuations for any additive observables:
$\langle \phi | \Delta \hat{A}^\dag \Delta \hat{A}|\phi \rangle = o(n^2)$,
where $\Delta \hat{A} = \hat{A} - \langle \phi | \hat{A} |\phi \rangle$.
That is, the relative fluctuations vanish in the macroscopic limit:
$\lim_{n \to 0} \langle \phi | \Delta \hat{A}^\dag \Delta \hat{A}|\phi \rangle^{1/2} / \langle \phi | \hat{A} |\phi \rangle = 0$.

Then, taking the contrapositive of the above statement,
we obtain a sufficient condition for a superposition of macroscopically distinct states:
a pure state $|\psi \rangle$ is a superposition of macroscopically distinct states
if $\langle \psi | \Delta \hat{A}^\dag \Delta \hat{A} |\psi \rangle = O(n^2)$
for an additive operator $\hat{A}$.
In other words, a superposition of macroscopically distinct states
has anomalous or macroscopic fluctuation for at least one additive observable.

\subsection{Index $p$: an indicator of superposition of macroscopically distinct states}\label{sec:index_p}

From the argument in the previous subsection,
we can judge whether a given state is a superposition of macroscopically distinct states
from the asymptotic behaviors of the fluctuations of additive observables.
To investigate the asymptotic behaviors,
we must assign states for each $n$ at a certain criterion.
We use a label $\nu$ to distinguish the assigned states for each $n$.
In AQC, for example, $\nu$ represents a problem instance of size $n$.
We collect all the assigned states to have a family
$F \equiv \{ | \psi_{n,\nu} \rangle \}_{n,\nu}$.

Then, we investigate asymptotic properties of the family.
We define the index $p$ for the family $F$ by
\begin{align}
\max_{\hat{A}(n)} \langle \Delta \hat{A}^\dag(n) \Delta \hat{A}(n) \rangle_{n,\nu}
= \overline{\Theta}(n^p),
\label{def:p}
\end{align}
where $\langle \cdots \rangle_{n,\nu} = \langle \psi_{n,\nu} | \cdots |\psi_{n,\nu} \rangle$,
and the maximum is taken over the additive operators $\hat{A}(n)$ in the system of size $n$.
The asymptotic notation $\overline{\Theta}$ is defined as follows \cite{Shimizu_etal:2013}.
Suppose a family of positive numbers $\{ f_\nu(n) \}_{n,\nu}$
that are generated by positive-valued functions $f_\nu(n)$ $(\nu = 1, 2, ...)$.
We say that $f_\nu(n) = \overline{\Theta}\bigl( g(n) \bigr)$ if and only if
$f_\nu(n) = \Theta\bigl( g(n) \bigr)$ for almost every $\nu$ in the family,
i.e., except for a vanishingly small fraction of $\nu$'s (for large $n$).
The big Theta notation is defined according to the convention in computer science \cite{Knuth:1976, Cormen_etal, NielsenChuang}:
$f(n) = \Theta\bigl( g(n) \bigr)$ if and only if $\gamma_1 g(n) \le f(n) \le \gamma_2 g(n)$
holds with positive constants $\gamma_1$ and $\gamma_2$ for $n \ge \exists n_0$.

We can show that $1 \le p \le 2$ if $p$ exists for $F$ \cite{Shimizu_etal:2013}.
It is clear that if $p = 2$ for a family $F$,
almost every state in $F$ is superpositions of macroscopically distinct states.
In this sense, the index $p$ is an indicator of superpositions of macroscopically distinct states.

We note that the index $p$ quantifies quantum entanglement
because superpositions of macroscopically distinct states are entangled states.
Furthermore it is known that $p$ is not correlated with bipartite entanglement (such as the entanglement entropy):
some states with $p = 2$ have small bipartite entanglement,
and some states with $p = 1$ have large bipartite entanglement \cite{Morimae_etal:2005, Sugita_etal:2005}.
Thus $p$ captures an aspect of entanglement that cannot be well quantified by bipartite entanglement.
See Refs.~\cite{Frowis_etal:2012, Yadin:2015, Tichy_etal:2016, Abad_etal:2016, Park_etal:2016, Kuwahara_etal:2017, Tatsuta_etal:2017}
for recent applications of the index $p$ and its relation to other quantities.

\subsection{Practical method of evaluating the index $p$}\label{sec:VCM}

From the definition of the index $p$, Eq.~(\ref{def:p}), it seems hard to calculate $p$ in general.
However, there is a method of efficiently evaluating $p$ \cite{Shimizu_etal:2013, Morimae_etal:2005, Sugita_etal:2005, Morimae_etal:2006}.

To explain the method, we consider a complete orthonormal set of local operators on $\mathcal{H}_\mathrm{loc}$ at site $l$---we write them as $\{ \hat{b}_\alpha(l) \}_{\alpha=0}^D$.
Here, $D = (\mathrm{dim}\mathcal{H}_\mathrm{loc})^2 - 1$
and $\hat{b}_0(l) = \hat{1}$ (identity operator on $\mathcal{H}_\mathrm{loc}$).
In a system composed of qubits, for example, $D = 3$,
and we may take $\hat{b}_\alpha(l)$'s ($\alpha = 1,2,3$) as the $x,y,z$ components of
the Pauli matrices $\hat{\sigma}_x(l), \hat{\sigma}_y(l), \hat{\sigma}_z(l)$ at site $l$.

In this method we evaluate the maximum eigenvalue of the variance-covariance matrix (VCM).
The VCM is defined by the following matrix elements:
\begin{align}
V_{l\alpha, l'\alpha'}(n,\nu) \equiv \langle \Delta \hat{b}^\dag_\alpha(l) \Delta \hat{b}_{\alpha'}(l') \rangle_{n,\nu},
\notag
\end{align}
where $\Delta \hat{b}^\dag_\alpha(l) = \hat{b}^\dag_\alpha(l) - \langle \hat{b}^\dag_\alpha(l) \rangle_{n,\nu}$.
Here, $l$ and $l'$ run over all the sites,
and $\alpha, \alpha' = 1, 2, ..., D$ ($\alpha=0$ and $\alpha'=0$ are not included),
so that the VCM is an $nD \times nD$ Hermitian matrix.
From the asymptotic behavior of the maximum eigenvalue $e_\mathrm{max}$ of the VCM,
we define another index $p_e$:
\begin{align}
e_\mathrm{max}(n,\nu) = \overline{\Theta}(n^{p_e - 1}).
\notag
\end{align}
Then we can show that $p = 2$ if and only if $p_e = 2$ \cite{Shimizu_etal:2013}.
Therefore, we can judge $p = 2$ by investigating whether $e_\mathrm{max}$ is asymptotically proportional to $n$.

Moreover, if $p = 2$, this method tells us which is the additive observable that macroscopically fluctuates \cite{Shimizu_etal:2013}.
We can construct it from the eigenvector $\bm{u}^\mathrm{max}$ corresponding to $e_\mathrm{max}$ as
$\hat{A} = \sum_{l=1}^n \sum_{\alpha=1}^D u^\mathrm{max}_{l \alpha} \hat{b}_\alpha(l)$.

\section{Conjecture on a necessary condition for quantum speedup in adiabatic quantum computation}\label{sec:conjecture}

In this section, we propose a conjecture on a necessary condition for the quantum speedup in AQC.
In Refs.~\cite{Ukena_etal:2004, Shimizu_etal:2013},
a conjecture was proposed for circuit-based quantum computation.
It states that a state with $p = 2$ is necessary for the quantum speedup.
Our expectation here is that both adiabatic and circuit-based quantum computation should use
the same features of quantum physics to achieve the quantum speedup,
although their operation principles are different.

Among several kinds of quantum speedup \cite{Ronnow_etal:2014},
we here consider the strong quantum speedup \cite{Papageorgiou_etal:2013}.
Our conjecture reads:
\begin{conjecture}
Let QAA be a quantum adiabatic algorithm that solves a problem.
If QAA achieves the strong quantum speedup,
then there exists a set $\mathsf{H}$ of infinitely many instances
such that a state with $p = 2$ appears during the computation of almost every instance in $\mathsf{H}$.
\end{conjecture}
Furthermore, if an algorithm uses only the ground states of $\hat{H}(s;n,\nu)$,
i.e., if the adiabatic approximation works well,
our conjecture yields a stronger form:
\begin{corollaryConjecture}
Let QAA be a quantum adiabatic algorithm that solves a problem
by using only the ground states $| \psi_{n,\nu}^0(s) \rangle$ of $\hat{H}(s;n,\nu)$.
If QAA achieves the strong quantum speedup,
then there exists a set $\mathsf{H}$ of infinitely many instances
such that $| \psi_{n,\nu}^0(s) \rangle$ at some $s$ has $p = 2$ for almost every $(n,\nu)$ in $\mathsf{H}$.
\end{corollaryConjecture}

Here we make a few remarks.

In Ref.~\cite{Shimizu_etal:2013}, problems are restricted to those not
in class BPP (bounded-error probabilistic polynomial time).
In our conjecture, we remove this restriction because at present there is no evidence for it.
In fact, for the Bernstein--Vazirani problem, which is in class BPP,
we obtain a result that supports our conjecture (Sec.~\ref{sec:BernsteinVazirani}).

One might think that our conjecture is a trivial extension of that in circuit-based quantum computation
since AQC is equivalent to circuit-based quantum computation up to polynomial overhead
\cite{Aharonov_etal:2007, Kempe_etal:2006, Mizel_etal:2007, Gosset_etal:2015}.
However, this is not the case for the following reasons.
(i) In the proofs of the equivalence,
mainly shown is the equivalence of the measurement output from the final states,
and the equivalence of (intermediate) states is not necessarily required.
Therefore, even if a state with $p=2$ appears in a circuit-based quantum algorithm,
it is not clear whether the same or similar state with $p=2$ appears in its equivalent algorithm of AQC.
(ii) If the quantum speedup is polynomial, the polynomial overhead can invalidate the speedup.
Therefore, even if a circuit-based quantum algorithm achieves the polynomial quantum speedup,
its equivalent algorithm of AQC does not necessarily achieve the speedup.
(iii) Some of the algorithms of AQC that achieve the quantum speedup were formulated without relying on the equivalence.
For such algorithms, validity of our conjecture is not trivial.
The algorithms investigated in Sec.~\ref{sec:examples} are examples of such ones.

As mentioned in Sec.~\ref{sec:index_p}, $p$ is defined for a family of states.
For the case of Corollary of Conjecture, we should take the family as
\begin{align}
F_{\mathsf H} = \bigl\{ | \psi_{n,\nu}^0(s_*) \rangle ~\big|~ \forall(n,\nu) \in \mathsf H, ~0 \le \exists s_* \le 1 \bigr\},
\notag
\end{align}
where $s_*$ may or may not depend on $(n,\nu)$.
In practice, $s_* = \mathrm{argmax}_{0\le s \le 1} e_\mathrm{max}(s;n,\nu)$ is useful.

It was pointed out \cite{Nest:2013}
that the quantum speedup can be achieved with vanishingly small entanglement.
If a problem can be solved by using an $n$-qubit AQC with states $\{ | \psi (s) \rangle \}_{0\le s \le 1}$,
the problem can be solved also by an $(n+1)$-qubit AQC with states
$\{| \Psi (s) \rangle = \sqrt{\varepsilon} | \psi (s) \rangle | 1 \rangle
+ \sqrt{1-\varepsilon} | 0^{\otimes n} \rangle | 0 \rangle \}_{0\le s \le 1}$
where $\varepsilon = 1/\mathrm{poly}(n)$.
In the latter AQC, any family of states has $p=1$ even if it achieves the quantum speedup.
However, because the parts essentially contributed to the computation are $\{ | \psi (s) \rangle \}$,
we circumvent this drawback by extracting the essential parts.
As one of the ways, we allow a state connected by a local operation when we evaluate the index $p$.
That is, we extend the definition (\ref{def:p}) of $p$ to
\begin{align}
\max_{\hat{\ell}_\alpha} \max_{\hat{A}(n)} \frac{\langle \psi_{n,\nu} | \hat{\ell}_\alpha^\dag
\Delta \hat{A}^\dag(n) \Delta \hat{A}(n) \hat{\ell}_\alpha | \psi_{n,\nu} \rangle}
{\langle \psi_{n,\nu} | \hat{\ell}_\alpha^\dag \hat{\ell}_\alpha | \psi_{n,\nu} \rangle}
= \overline{\Theta}(n^p),
\label{extendedDef:p}
\end{align}
where
$\Delta \hat{A} = \hat{A} - \langle \psi_{n,\nu} | \hat{\ell}_\alpha^\dag \hat{A}(n) \hat{\ell}_\alpha | \psi_{n,\nu} \rangle / \langle \psi_{n,\nu} | \hat{\ell}_\alpha^\dag \hat{\ell}_\alpha | \psi_{n,\nu} \rangle$,
and $\hat{\ell}_\alpha$ is a Kraus operator on $O(1)$ sites (qubits),
which represents a local operation.
The leftmost maximum in Eq.~(\ref{extendedDef:p}) is taken over all the local operations
that satisfy $\langle \psi_{n,\nu} | \hat{\ell}_\alpha^\dag \hat{\ell}_\alpha | \psi_{n,\nu} \rangle \ge 1/\mathrm{poly}(n)$.
In the next section, however, we use the original definition (\ref{def:p})
because we can evaluate $p$ by Eq.~(\ref{def:p}) for the states there.

\section{Tests for conjecture}\label{sec:examples}

Our conjecture for AQC as well as the conjecture for circuit-based quantum computation \cite{Shimizu_etal:2013}
have not been proved for \textit{general} problems at present.
In this section, we investigate its correctness in \textit{specific} problems as the first tests for our conjecture.
We have five examples of AQC algorithms:
quantum adiabatic versions of the algorithms of
Grover, Deutsch--Jozsa, Bernstein--Vazirani, and Simon and an algorithm for the glued-trees problem.
All of these algorithms are known to achieve the strong quantum speedup or not.
We show the summary of the results in Table \ref{table:results}.
All the results are consistent with and thus support our conjecture.

Throughout this section, we assume systems of qubits.
We use the following notation.
The Pauli matrices at site $l$ are $\hat{\sigma}_x(l), \hat{\sigma}_y(l), \hat{\sigma}_z(l)$.
The eigenstate of $\hat{\sigma}_z(l)$ corresponding to the eigenvalue of $+1$ is $| 0 \rangle_l$,
and that of $-1$ is $| 1 \rangle_l$.
The eigenstates of $\hat{\sigma}_x(l)$ corresponding to the eigenvalues of $\pm 1$
are $| \pm \rangle_l = (| 0 \rangle_l \pm | 1 \rangle_l) / \sqrt{2}$.
The computational basis in the $n$-qubit system is composed of
$| w \rangle = \bigotimes_{l=1}^{n} | w_l \rangle_l$ with $w \in \{0,1\}^n = \{ 0, 1, ... , N - 1 \}$.
Here, $w_l \in \{0,1\}$ is the $l$th-bit value of $w$, and $N = 2^n$.

\subsection{Adiabatic Grover algorithm}\label{sec:Grover}

The Grover algorithm is a quantum algorithm of searching solutions
in unstructured databases of $N = 2^n$ entries.
It was originally formulated in a circuit-based quantum computation \cite{Grover:1997}
and was proved to achieve the quadratic quantum speedup
\cite{Grover:1997, Bennett_etal:1997, NielsenChuang, Mermin}.

An AQC version of the Grover algorithm was proposed in Ref~\cite{Farhi_etal:2000}.
It was improved in Ref~\cite{Roland_etal:2002}
to achieve the quadratic quantum speedup \cite{Roland_etal:2002, Jansen_etal:2007, Albash_etal:2016}.

Here, we show that our conjecture is correct for the adiabatic Grover algorithm of Ref~\cite{Roland_etal:2002}.
For simplicity we assume single-solution cases.

The input of Grover's search problem is a function $f$ of $n$-bit strings, $f: \{0,1\}^n \to \{0,1\}$,
which satisfies $f(w) = 1$ only for the solution $w = w^* \in \{0,1\}^n$
and $f(w) = 0$ for $\forall w \in \{0,1\}^n$ except for $w^*$.
The task is to find the solution $w^*$.
An instance for fixed $n$ is identified with $w^*$---we therefore use $w^*$ instead of $\nu$.

For the adiabatic Grover algorithm \cite{Roland_etal:2002},
the problem Hamiltonian is given by $\hat{H}_p(n,w^*) = 1 - | w^* \rangle \langle w^* |$,
where $| w^* \rangle$ is one of the computational basis states corresponding to the solution $w^*$.
And the driver Hamiltonian is given by
\begin{align}
\hat{H_d}(n) &= 1 - | \varphi \rangle \langle \varphi |,
\label{Hd_Grover}
\end{align}
where $| \varphi \rangle$ is the uniform superposition of the computational basis,
which is equivalent to the product state of $| + \rangle_l$'s:
\begin{align}
| \varphi \rangle \equiv \frac{1}{\sqrt{N}} \sum_{w=0}^{N-1} | w \rangle
= \bigotimes_{l=1}^n | + \rangle_l.
\label{plus_x}
\end{align}

In this algorithm, $s(t)$ is tuned to satisfy a time-local adiabatic condition \cite{Roland_etal:2002},
so that the system is in the ground state $| \psi_{n,w^*}^0(s) \rangle$ of $\hat{H}(s;n,w^*)$ at each $s$.
We can show that the ground state lies in the subspace spanned by $| w^* \rangle$ and $| \varphi \rangle$
\cite{Roland_etal:2002, Albash_etal:2016}.
We thus easily obtain $| \psi_{n,w^*}^0(s) \rangle$ as
\begin{align}
| \psi_{n,w^*}^0(s) \rangle &= a_s | w^* \rangle + b_s | \varphi \rangle,
\label{gs_Grover}
\end{align}
where
\begin{align}
&a_s = \sin \frac{\theta_s}{2} - \cos \frac{\theta_s}{2} \tan \frac{\theta_0}{2},
~~~
b_s = \cos \frac{\theta_s}{2} \bigg/ \cos \frac{\theta_0}{2},
\notag
\\
&\sin \theta_s = \frac{2 (1 - s) \sqrt{N-1}}{\Delta E(s) N},
\notag
\\
&\cos \theta_s = - \frac{1}{\Delta E(s)} \left[ 1 - 2 (1 - s) \frac{N-1}{N} \right],
\notag
\\
&\Delta E(s) = \sqrt{1 - 4 s (1-s) \frac{N-1}{N}}.
\notag
\end{align}
We note that $\Delta E(s)$ is the energy gap between the ground and first-excited states
and that the gap closes at $s=1/2$ in the infinite-size limit.

Now we show that our conjecture is correct for the adiabatic Grover algorithm.
Because the algorithm uses only the ground states, we investigate Corollary of Conjecture.
We choose the set $\mathsf{H}$ in our conjecture as the collection of all the instances:
$\mathsf{H} = \bigl\{ (n,w^*) ~\big|~ \forall w^* \in \{0,1\}^n , n = 1,2,... \bigr\}$.
Then we take the family of states $F_{\mathsf H}(s)$ as the ground states at fixed $s$:
\begin{align}
F_{\mathsf H}(s) = \bigl\{ | \psi_{n,w^*}^0(s) \rangle \bigr\}_{(n,w^*) \in \mathsf H}.
\notag
\end{align}
To $p = 2$, it is sufficient to find one additive observable whose fluctuation is macroscopically large.
We here choose $\hat{M}_x \equiv \sum_{l=1}^n \hat{\sigma}_x(l)$ as the additive observable
(this choice is the same as that in Ref.~\cite{Shimizu_etal:2013} for the circuit-based Grover algorithm).

By using Eqs.~(\ref{gs_Grover}) and (\ref{plus_x}),
we can calculate the expectation values of $\hat{M}_x$ and $\hat{M}_x^2$ for $| \psi_{n,w^*}^0(s) \rangle$ as
\begin{align}
\langle \psi_{n,w^*}^0(s) | \hat{M}_x | \psi_{n,w^*}^0(s) \rangle &= b_s^2 n + \frac{2 a_s b_s n}{\sqrt{N}},
\notag
\\
\langle \psi_{n,w^*}^0(s) | \hat{M}_x^2 | \psi_{n,w^*}^0(s) \rangle
&= b_s^2 n^2 + a_s^2 n + \frac{2 a_s b_s n^2}{\sqrt{N}}.
\notag
\end{align}
We thus obtain the fluctuation as
\begin{align}
&\langle \psi_{n,w^*}^0(s) | \Delta \hat{M}_x^2 | \psi_{n,w^*}^0(s) \rangle
\label{fluctuation_Grover}
\\
&= b_s^2 (1 - b_s^2) n^2 + a_s^2 n + 2 a_s b_s (1 - 2b_s^2) \frac{n^2}{\sqrt{N}} - 4 a_s^2 b_s^2 \frac{n^2}{N}.
\notag
\end{align}
It is possible only for the first term to be $\Theta(n^2)$.
We can evaluate the prefactor of this term as
\begin{align}
b_s^2 (1 - b_s^2) =
\begin{cases}
1/4 + O(N^{-1}) & s = 1/2
\\
O(N^{-1}) & s \neq 1/2.
\end{cases}
\label{prefactor_Grover}
\end{align}
Equations~(\ref{fluctuation_Grover}) and (\ref{prefactor_Grover}) lead to the result
that $p = 2$ for the family $F_{\mathsf H}(s=1/2)$ at the gap closing point for every instance.
We thus conclude that Corollary of Conjecture is correct for the adiabatic Grover algorithm.

We may understand this result as follows.
As shown in Eq.~(\ref{gs_Grover}), the ground state can be expressed as
a superposition of the two states, $| w^* \rangle$ and $| \varphi \rangle$.
These two states are macroscopically distinct
because $\langle w^* | \hat{M}_x | w^* \rangle = 0$ and $\langle \varphi | \hat{M}_x | \varphi \rangle = n$.
For large $n$, the amplitudes, $a_s$ and $b_s$, in Eq.~(\ref{gs_Grover}) are approximated
within errors of $O(N^{-1/2})$ as
\begin{align}
a_s \simeq
\begin{cases}
0 & s < 1/2
\\
1/\sqrt{2} & s = 1/2
\\
1 & s > 1/2
\end{cases}
\notag
\end{align}
and $b_s \simeq \sqrt{1 - a_s^2}$.
This indicates that a transition between the macroscopically distinct states
(from the initial state $| \varphi \rangle$ to the final state $| w^* \rangle$)
occurs at the gap closing point ($s=1/2$)
and that the superposition of these states appears at the transition point.

\subsection{Adiabatic Deutsch--Jozsa algorithm}\label{sec:DeutschJozsa}

The Deutsch--Jozsa algorithm is a quantum algorithm to solve the following problem
\cite{NielsenChuang, Mermin, Deutsch_etal:1992}.
The input of the problem is a function $f$ of $n$-bit strings, $f: \{0,1\}^n \to \{0,1\}$.
The function is promised to be either constant or balanced:
$f$ is constant if $f(w) = 0$ for any $w \in \{0,1\}^n$ or if $f(w) = 1$ for any $w \in \{0,1\}^n$,
whereas $f$ is balanced if $f(w) = 0$ for $w$ in a half of $\{0,1\}^n$
and $f(w) = 1$ for $w$ in the other half.
The problem is to judge which type a given function $f$ is.
An instance of the problem for fixed $n$ is identified with $f$---we therefore use $f$ instead of $\nu$.

The original Deutsch--Jozsa algorithm was proposed in the circuit-based quantum computation \cite{Deutsch_etal:1992}.
It was shown that the quantum algorithm solves the problem in a single query of $f$,
i.e., its computational time is $O(1)$ \cite{Cleve_etal:1998}.
Note that, however, a classical bounded-error probabilistic algorithm can also solve the problem
in $n$-independent times of $f$-queries
(although classical deterministic algorithms requires $O(2^n)$ computational time).
Hence, the Deutsch--Jozsa algorithm does not achieve the quantum speedup.

AQC versions of the algorithm were proposed in Refs.~\cite{Sarandy_etal:2005, Wei_etal:2006}.
In these algorithms the energy gap is $O(1)$ between the ground and first-excited states of $\hat{H}(s;n,f)$.
From this, it follows that they require $O(1)$ computational time to satisfy the adiabatic condition.
This also implies that they use only the ground states.

We investigate whether Corollary of Conjecture is correct in this problem.
We here focus on the algorithm proposed in Ref.~\cite{Wei_etal:2006}.
As we show below, only states with $p = 1$ appear in this algorithm.
This implies that the contrapositive of our conjecture is correct for this algorithm:
if states with $p = 2$ never appear during the computation, the algorithm will not achieve the quantum speedup.

For the adiabatic Deutsch--Jozsa algorithm in Ref.~\cite{Wei_etal:2006},
the driver Hamiltonian $\hat{H_d}(n)$ is the same as that of the adiabatic Grover algorithm, Eq.~(\ref{Hd_Grover}) with $| \varphi \rangle$ in Eq.~(\ref{plus_x}).
The problem Hamiltonian is $\hat{H}_p(n,f) = 1 - | \beta_f \rangle \langle \beta_f |$,
where the final state $| \beta_f \rangle$ is chosen as
\begin{align}
| \beta_f \rangle &= \frac{\mu_f}{\sqrt{N/2}} \sum_{w' = 0}^{N/2 - 1} | 2w' \rangle
+ \frac{1 - \mu_f}{\sqrt{N/2}} \sum_{w' = 0}^{N/2 - 1} | 2w' + 1 \rangle,
\label{final_state_DeutschJozsa}
\\
\mu_f &\equiv \frac{1}{N} \left| \sum_{w=0}^{N-1} (-1)^{f(w)} \right| =
\begin{cases}
1 & f:\mathrm{constant}
\\
0 & f:\mathrm{balanced},
\end{cases}
\notag
\end{align}
with $N = 2^n$.
The above equation means that $| \beta_f \rangle$ is the uniform superposition of all the
even (odd) index states in the computational basis
if $f$ is constant (balanced)---hence we can get an answer
by measuring the final state $| \beta_f \rangle$ in the computational basis.
We note that only a particular single qubit, say the first qubit ($l=1$),
has the information on whether a state in the computational basis is an even index state or odd one.
Hence we can rewrite Eq.~(\ref{final_state_DeutschJozsa}) as
\begin{align}
| \beta_f \rangle
&= \Bigl[ \mu_f | 0 \rangle_1 + (1 - \mu_f) | 1 \rangle_1 \Bigr]
\otimes \frac{1}{\sqrt{N/2}} \sum_{w' = 0}^{N/2 - 1} | w' \rangle
\notag
\\
&= \Bigl[ \mu_f | 0 \rangle_1 + (1 - \mu_f) | 1 \rangle_1 \Bigr] \bigotimes_{l = 2}^n | + \rangle_l.
\label{beta_f}
\end{align}

We can easily show that the ground state of $\hat{H}(s;n,f)$ lies in the subspace
spanned by $| \beta_f \rangle$ and $| \varphi \rangle$.
We thus obtain the ground state $| \psi_{n,f}^0(s) \rangle$ as
\begin{align}
| \psi_{n,f}^0(s) \rangle
= \sin \frac{\theta_s}{2} | \beta_f \rangle + \cos \frac{\theta_s}{2} | \beta_f^\perp \rangle,
\label{gs_DeutschJozsa}
\end{align}
where
\begin{align}
| \beta_f^\perp \rangle = \sqrt{2} | \varphi \rangle - | \beta_f \rangle.
\label{beta_f_perp}
\end{align}
The concrete form of $\theta_s$ is not important here.
By substituting Eqs.~(\ref{beta_f_perp}), (\ref{beta_f}), and (\ref{plus_x}) into Eq.~(\ref{gs_DeutschJozsa}),
we obtain
\begin{align}
| \psi_{n,f}^0(s) \rangle
= \Bigl[ c_{0,f}(s) | 0 \rangle_1 + c_{1,f}(s) | 1 \rangle_1 \Bigr] \bigotimes_{l = 2}^n | + \rangle_l,
\end{align}
where
$c_{0,f}(s) = (1 - \mu_f) \cos \frac{\theta_s}{2} + \mu_f \sin \frac{\theta_s}{2}$,
and
$c_{1,f}(s) = (1 - \mu_f) \sin \frac{\theta_s}{2} + \mu_f \cos \frac{\theta_s}{2}$.
Therefore the ground state $| \psi_{n,f}^0(s) \rangle$ at any $s$ is a product state for every instance.

This result means that $p = 1$ for the family of the ground states at any $s$ and for any set of instances,
since (any families of) product states have $p = 1$ \cite{Shimizu_etal:2013}.
We therefore conclude that the contrapositive of our conjecture is correct
for the adiabatic Deutsch--Jozsa algorithm.

\subsection{Adiabatic Bernstein--Vazirani algorithm}\label{sec:BernsteinVazirani}

The Bernstein--Vazirani algorithm is a quantum algorithm to solve the following problem \cite{Bernstein_etal:1997, Mermin}.
The input of the problem is a function $f_a$ of $n$-bit strings, $f_a: \{0,1\}^n \to \{0,1\}$,
which is given by the bitwise inner product with a hidden $n$-bit string $a \in \{0,1\}^n$ modulo 2:
\begin{align}
f_a(w) = \left( \sum_{l=1}^n a_l w_l \right) ~~ \mathrm{mod}~2,
\notag
\end{align}
where $a_l$ ($w_l$) is the $l$th-bit value of $a$ ($w$).
The task is to find the value of $a$.
An instance of the problem for fixed $n$ is identified with $a$---we therefore use $a$ instead of $\nu$.

The original Bernstein--Vazirani algorithm was proposed
in a context of quantum Turing machine \cite{Bernstein_etal:1997}.
It was shown that the quantum algorithm solves the problem in $O(1)$ queries of $f_a$
while classical algorithms require $O(n)$ queries---hence the polynomial quantum speedup is achieved.

An AQC version of the algorithm was proposed in Ref~\cite{Hen:2014b}.
It was shown that this algorithm requires $O(1)$ computational time,
so that it also achieves the polynomial quantum speedup.

Before investigating our conjecture, we briefly review this adiabatic algorithm.
This algorithm uses a system of $n + 1$ qubits for instances of size $n$.
The problem Hamiltonian is given by
\begin{align}
\hat{H}_p (n,a) = \frac{1}{2} \sum_{w = 0}^{N - 1} | w \rangle \langle w |
\otimes \bigl[ \hat{1} + (-1)^{f_a(w)} \hat{\sigma}_z(n+1) \bigr],
\notag
\end{align}
where $N = 2^n$.
And the driver Hamiltonian is given by
\begin{align}
\hat{H}_d(n) &= \frac{1}{2} \left( \bigotimes_{l=1}^n \hat{1} \right)
\otimes \bigl[ \hat{1} - \hat{\sigma}_x(n+1) \bigr].
\notag
\end{align}
Obviously, the ground states are $N$-fold degenerate for both $\hat{H}_p$ and $\hat{H}_d$.
We choose the initial state as the following one of the degenerate ground states of $\hat{H}_d$:
\begin{align}
| \psi_{n,a}^0(0) \rangle 
= \bigotimes_{l=1}^{n+1} | + \rangle_l.
\label{initialState_BV}
\end{align}
Then, after we slowly vary $s$ from $s=0$ to $s=1$ in $\hat{H}(s;n,a)$ of Eq.~(\ref{Hs}),
we can adiabatically connect this initial state to the following one of the ground states of $\hat{H}_p$:
\begin{align}
| \psi_{n,a}^0(1) \rangle = \frac{1}{\sqrt{N}} \sum_{w=0}^{N-1} | w \rangle \otimes | f_a(w) \rangle_{n+1}.
\label{gs_BV_s1}
\end{align}
The time needed for this adiabatic evolution is $O(1)$
because the energy gap between the ground states and the first excited states is independent of $n$ \cite{Hen:2014b}.
We then measure $\hat{\sigma}_x(n+1)$ on $| \psi_{n,a}^0(1) \rangle$.
We obtain the outcome of $-1$ with probability $1/2$ to have a post-measurement state,
\begin{align}
| \psi_{n,a}^- \rangle = \bigotimes_{l=1}^n \bigl( | 0 \rangle_l + (-1)^{a_l}| 1 \rangle_l \bigr)
\otimes | - \rangle_{n+1}.
\label{finalState_BV}
\end{align}
Since this state corresponds to the configuration of $a$ in the $x$ basis,
we obtain the solution $a$ by performing a measurement of $\{ \hat{\sigma}_x(l) \}_{l=1}^n$ on this state.
In $n$-independent runs of this procedure, we can make exponentially small
the probability to fail to obtain $-1$ in the measurement of $\hat{\sigma}_x(n+1)$.
Therefore the computational time of this adiabatic algorithm is $O(1)$.

Now we show that our conjecture is correct for this algorithm.
We take the set $\mathsf{H}$ as the collection of all the instances:
$\mathsf{H} = \bigl\{ (n,a) ~\big|~ \forall a \in \{0,1\}^n, n = 1, 2, ... \bigr\}$.
Then we find that $p = 2$ for the family of the states at the end of the adiabatic evolution:
\begin{align}
F_{\mathsf H} = \bigl\{ | \psi_{n,a}^0(1) \rangle \bigr\}_{(n,a) \in \mathsf H},
\label{family_BV}
\end{align}
where $| \psi_{n,a}^0(1) \rangle$ is given by Eq.~(\ref{gs_BV_s1}).
In the following, we show this result by using the method of the VCM (described in Sec.~\ref{sec:VCM}).

To this end, we first note the following exact equality
for the maximum eigenvalue $e_\mathrm{max}(n,a)$ of the VCM for $| \psi_{n,a}^0(1) \rangle$:
\begin{align}
e_\mathrm{max}(n,a) =
\begin{cases}
2 & a = 0
\\
1 + \sum\limits_{l=1}^n a_l & 1 \le a \le N - 1.
\end{cases}
\label{e_max_BV}
\end{align}
See Appendix \ref{appendix:derivation_of_e_max_BV} for the derivation of this equality.
Next, suppose that $a$ is randomly chosen from $\{0,1\}^n$ according to the uniform distribution.
Then, Eq.~(\ref{e_max_BV}) implies that $e_\mathrm{max}(n,a)$ follows the binomial distribution $B(n,1/2)$.
For large $n$, because $B(n,1/2)$ is well approximated by the normal distribution $\mathcal{N}(n/2, n/4)$,
$| e_\mathrm{max}(n,a) - n/2 | \lesssim \sqrt{n}$ holds for almost every instance.
This leads to $e_\mathrm{max}(n,a) = \overline{\Theta}(n)$ for the states in $F_{\mathsf H}$ of Eq.~(\ref{family_BV})
(for rigorous estimation, see Appendix~\ref{appendix:estimation_of_e_max}).
We thus show that $p_e = 2$ and $p = 2$ for $F_{\mathsf H}$.
Therefore, we conclude that Corollary of Conjecture is correct for the adiabatic Bernstein--Vazirani
algorithm.

We may understand this result as follows.
As shown in Appendix \ref{appendix:derivation_of_e_max_BV},
we can rewrite $| \psi_{n,a}^0(1) \rangle$ as
\begin{align}
&| \psi_{n,a}^0(1) \rangle
= \frac{1}{\sqrt{2}} \bigl( | \psi_{n,a}^0(0) \rangle + | \psi_{n,a}^- \rangle \bigr),
\label{gs_BV_s1_rewrite}
\end{align}
where $| \psi_{n,a}^0(0) \rangle$ is the initial state [Eq.~(\ref{initialState_BV})]
and $| \psi_{n,a}^- \rangle$ is the configuration of the solution [Eq.~(\ref{finalState_BV})].
We note that both $| \psi_{n,a}^0(0) \rangle$ and $| \psi_{n,a}^- \rangle$ are eigenstates
of an additive operator $\hat{A} = \sum_{l=1}^n a_l \hat{\sigma}_x(l) + \hat{\sigma}_x(n+1)$
and the difference of their eigenvalues is $2 \sum_{l=1}^n a_l + 2$.
Since $\sum_{l=1}^n a_l = \overline{\Theta}(n)$ as argued above,
These two states are macroscopically distinct for almost every $a$.
Equation~(\ref{gs_BV_s1_rewrite}) indicates that $| \psi_{n,a}^0(1) \rangle$ is a superposition of these two states.

\subsection{Adiabatic Simon algorithm}\label{sec:Simon}

The Simon algorithm is a quantum algorithm of solving the following problem \cite{Simon:1997, Mermin}.
The input of the problem is a function $g$ of $n$-bit strings, $g: \{0,1\}^n \to \{0,1\}^{n-1}$,
which is characterized by a hidden $n$-bit string $a \in \{0,1\}^n - \{ 0 \}$.
The function $g$ is promised to satisfy $g(w) = g(w')$
if and only if $w \oplus a= w'$ (except for the trivial case of $w = w'$),
where the symbol $\oplus$ is the bitwise-xor operation.
The task is to find the value of $a$.
An instance of the problem for fixed $n$ is identified with $a$---we therefore use $a$ instead of $\nu$.

The original Simon algorithm was proposed in Ref.~\cite{Simon:1997}.
It was shown that the quantum algorithm solve the problem in $O(n)$ queries of $g$
while classical algorithms require exponentially many queries---hence the exponential quantum speedup
is achieved.

An AQC version of the algorithm was proposed in Ref.~\cite{Hen:2014b}.
It was shown that this algorithm also achieves the exponential quantum speedup.

Before investigating our conjecture, we briefly review this adiabatic algorithm.
This algorithm uses a composite system of two subsystems---for instances of size $n$,
the first subsystem consists of $n$ qubits and the second consists of $n - 1$ qubits.
The problem Hamiltonian is given by
\begin{align}
\hat{H}_p (n,a) = \frac{1}{2} \sum_{w = 0}^{N - 1} | w \rangle \langle w |
\otimes \sum_{l_2 = 1}^{n-1} \bigl[ \hat{1} + (-1)^{g_{l_2}(w)} \hat{\sigma}_z(l_2) \bigr],
\notag
\end{align}
where $N = 2^n$, and $g_{l_2}(w)$ is the $l_2$th-bit value of $g(w)$.
The driver Hamiltonian is given by
\begin{align}
\hat{H}_d(n) &= \frac{1}{2} \left( \bigotimes_{l_1=1}^n \hat{1} \right)
\otimes \sum_{l_2 = 1}^{n-1} \bigl[ \hat{1} - \hat{\sigma}_x(l_2) \bigr].
\notag
\end{align}
The ground states are $N$-fold degenerate for both $\hat{H}_p$ and $\hat{H}_d$.
Similarly to the case of the adiabatic Bernstein--Vazirani algorithm,
we choose the initial state as the following one of the degenerate ground states of $\hat{H}_d$:
$| \psi_{n,a}^0(0) \rangle = \bigotimes_{l_1=1}^{n} | + \rangle_{l_1} \bigotimes_{l_2=1}^{n-1} | + \rangle_{l_2}$.
Then, by slowly varying $s$ from $s=0$ to $s=1$ in $\hat{H}(s;n,a)$ of Eq.~(\ref{Hs}),
we can adiabatically connect this initial state to the following one of the ground states of $\hat{H}_p$:
$| \psi_{n,a}^0(1) \rangle = (1 / \sqrt{N}) \sum_{w=0}^{N-1} | w \rangle \otimes | g(w) \rangle$.
The time needed for this adiabatic evolution is $O(1)$
because the energy gap between the ground states and the first excited states is independent of $n$ \cite{Hen:2014b}.
After the adiabatic evolution, we perform the computational-basis measurement of the second subsystem.
We obtain an outcome $g(w^*)$ to have a post-measurement state,
\begin{align}
| \psi_{n,a,w^*} \rangle
= \frac{1}{\sqrt{2}} \bigl( | w^* \rangle + | w^* \oplus a \rangle \bigr) \otimes | g(w^*) \rangle.
\label{state_Simon}
\end{align}
We subsequently perform the $x$-basis measurement of the first subsystem
to obtain an outcome $x^* \in \{0,1\}^n$, which is orthogonal to $a$ in the bitwise inner product modulo 2.

In the algorithm we perform $O(n)$ runs of the above procedure
(initial-state preparation, adiabatic evolution, and two measurements).
Then, almost certainly we will obtain sufficient information to determine the value of $a$.
Because the computational time in a single run is $O(1)$,
overall time of the algorithm is $O(n)$---thus it achieves the exponential quantum speedup.

Now we show that our conjecture is correct for this algorithm.
We investigate Conjecture since it uses not only the ground states but also the post-measurement states.
We take the set $\mathsf{H}$ as the collection of all the instances:
$\mathsf{H} = \bigl\{ (n,a) ~\big|~ \forall a \in \{0,1\}^n, n = 1, 2, ... \bigr\}$.
Then we find that $p = 2$ for the family of the post-measurement states:
\begin{align}
F_{\mathsf H} = \bigl\{ | \psi_{n,a,w^*} \rangle \bigr\}_{(n,a) \in \mathsf H},
\label{family_Simon}
\end{align}
where $| \psi_{n,a,w^*} \rangle$ is given by Eq.~(\ref{state_Simon}),
and $w^*$ (determined by a random outcome) is arbitrary in $\{0,1\}^n$.
In the following, we show this result by using the method of the VCM (described in Sec.~\ref{sec:VCM}).

To this end, we note the following exact equality
for the maximum eigenvalue $e_\mathrm{max}(n,a)$ of the VCM for $| \psi_{n,a,w^*} \rangle$:
\begin{align}
e_\mathrm{max}(n,a) =
\begin{cases}
2 & a = 2^{l-1}~(l = 1, 2, ..., n)
\\
\sum\limits_{l=1}^n a_l & \mathrm{otherwise}.
\end{cases}
\label{e_max_Simon}
\end{align}
This holds independently of $w^*$.
(Note that the case of $a=0$ is not included in Simon's problem.)
See Appendix \ref{appendix:derivation_of_e_max_Simon} for the derivation of Eq.~(\ref{e_max_Simon}).
Hence, in the same way as in Sec.~\ref{sec:BernsteinVazirani},
we can show that $e_\mathrm{max}(n,a)$ follows the binomial distribution
and that for large $n$, $| e_\mathrm{max}(n,a) - n/2 | \lesssim \sqrt{n}$ holds for almost every instance.
This leads to $e_\mathrm{max}(n,a) = \overline{\Theta}(n)$ for the states in $F_{\mathsf H}$ of Eq.~(\ref{family_Simon})
(for rigorous estimation, see Appendix~\ref{appendix:estimation_of_e_max}).
We thus show that $p_e = 2$ and $p = 2$ for $F_{\mathsf H}$.
Therefore, we conclude that our conjecture is correct for the adiabatic Simon algorithm.

We may understand this result as follows.
Equation (\ref{state_Simon}) indicates that $| \psi_{n,a,w^*} \rangle$ is a superposition of
$| w^* \rangle \otimes | g(w^*) \rangle$ and  $| w^* \oplus a \rangle \otimes | g(w^*) \rangle$.
These two states are eigenstates of an additive operator
$\hat{A} = \sum_{l_1 = 1}^n (-1)^{w^*_{l_1}} a_{l_1} \hat{\sigma}_z(l_1)$,
and the difference of their eigenvalues is $2 \sum_{l_1 = 1}^n a_{l_1}$.
Since $\sum_{l=1}^n a_l = \overline{\Theta}(n)$ as argued above,
$| w^* \rangle \otimes | g(w^*) \rangle$ and  $| w^* \oplus a \rangle \otimes | g(w^*) \rangle$
are macroscopically distinct for almost every $a$.
Therefore, $| \psi_{n,a,w^*} \rangle$ is a superposition of these two macroscopically distinct states.

\subsection{Glued-trees problem}\label{sec:gluedTrees}

\begin{figure}[t]
\begin{center}
\includegraphics[width=\linewidth]{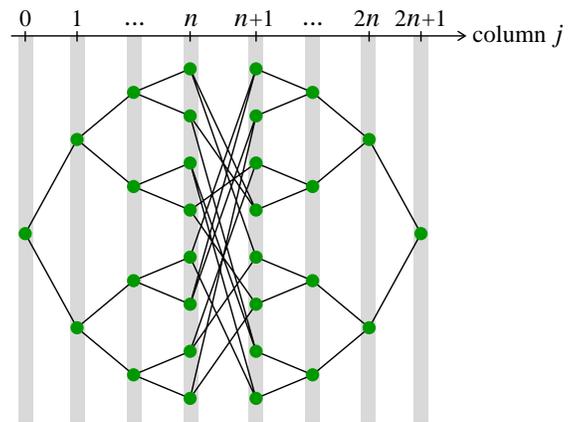}
\end{center}
\caption{
Glued-trees graph for an instance of $n=3$.
}
\label{fig:gluedTrees}
\end{figure}

The glued-trees problem is as follows \cite{Somma_etal:2012, Childs_etal:2003}.
We consider two perfect binary trees of height $n$, as shown in Fig.~\ref{fig:gluedTrees}.
The total number of the vertices is $M = 2^{n+2} - 2$.
Each vertex has a unique name randomly chosen from $2n$-bit strings---this implies that
the number of possible names is $N = 2^{2n}$ and is much larger than $M$.
The trees are randomly glued:
the leaves are connected by a random cycle that alternates between the leaves of the two trees.
We are given an oracle that outputs the names of the adjacent vertices on any input vertex name.
Then the problem is, given the name of the left root, to find the name of the right root.
An instance (denoted by $\nu$) for fixed $n$ is identified with a random cycle and a list of random names.

This problem was proposed in Ref.~\cite{Childs_etal:2003} in a context of quantum walk.
It was shown that a quantum-walk algorithm solves it in polynomial computational time,
while classical algorithms require at least subexponential number of oracle calls.
Hence the quantum algorithm achieves the exponential quantum speedup.

An AQC version of the algorithm was proposed in Ref.~\cite{Somma_etal:2012}.
It was shown that this adiabatic algorithm also achieves the exponential quantum speedup.

Before investigating our conjecture in this algorithm, we briefly review it.
This algorithm uses a system of $2n$ qubits for instances of problem size $n$
to express vertex names of $2n$-bit strings.
The total Hamiltonian is different from that in Eq.~(\ref{Hs}).
It is composed of three Hamiltonians:
\begin{align}
\hat{H}(s;n,\nu) = (1 - s) \alpha \hat{H}_\mathrm{ini} + s \alpha \hat{H}_\mathrm{fin} - s (1 - s) \hat{H}_\mathrm{ora},
\label{Hs_gluedTrees}
\end{align}
where the parameter $\alpha$ is a constant satisfying $0 < \alpha < 1/2$.
In the below, we define individual Hamiltonians.

Let $w(v) \in \{0,1\}^{2n}$ be the name of a vertex $v$.
We define the initial and final Hamiltonians respectively as
$\hat{H}_\mathrm{ini} = - | w(0) \rangle \langle w(0) |$
and  $\hat{H}_\mathrm{fin} = - | w(M-1) \rangle \langle w(M-1) |$.
Here, $v=0$ and $v=M-1$ represent the left and right roots, respectively.
To define the oracle Hamiltonian $\hat{H}_\mathrm{ora}$, we consider the following orthonormal states:
$| \mathrm{col}(j) \rangle = (1/\sqrt{N_j})\sum_{v \in V_j} | w(v) \rangle$.
Here, $j=0,1,...,2n+1$, and $V_j$ is the set of the vertices belonging to the $j$th column
(see Fig.~\ref{fig:gluedTrees}), and $N_j = \# V_j$
($N_j = 2^j$ for $0 \le j \le n$ and $N_j = 2^{2n+1 - j}$ for $n+1 \le j \le 2n+1$).
Note that $| \mathrm{col}(0) \rangle = | w(0) \rangle$ and $| \mathrm{col}(2n+1) \rangle = | w(M-1) \rangle$.
In the subspace spanned by $\{ | \mathrm{col}(j) \rangle \}_j$, we define the oracle Hamiltonian as
an Hermitian matrix with $\langle \mathrm{col}(j) | \hat{H}_\mathrm{ora} | \mathrm{col}(j') \rangle = 0$ for $j' \neq j \pm 1$, and
\begin{align}
\langle \mathrm{col}(j) | \hat{H}_\mathrm{ora} | \mathrm{col}(j+1) \rangle =
\begin{cases}
\sqrt{2} & j=n
\\
1 & \mathrm{otherwise}.
\end{cases}
\notag
\end{align}
It was shown that there are two crossing points, $s = s_c = \alpha / \sqrt{2}$ and $s = 1 - s_c$,
for $\hat{H}(s;n,\nu)$ in Eq.~(\ref{Hs_gluedTrees}) \cite{Somma_etal:2012}.
At these points, the energy gap between the ground and first excited states
becomes exponentially small as $n$ increases,
and the energy levels cross for the infinite-size limit.

The quantum adiabatic algorithm works as follows \cite{Somma_etal:2012}.
We initialize the state in $| w(0) \rangle$, the ground state of $\hat{H}(s;n,\nu) = \hat{H}_\mathrm{ini}$,
and increase $s$ at a constant rate of $\dot{s}(t) = O(n^{-6})$.
Then, the state adiabatically evolves in the ground state $| \psi^0_{n,\nu}(s) \rangle$
until a transition to the first excited state occurs at around $s = s_c$.
After the first transition, the state adiabatically evolves in the first excited state $| \psi^1_{n,\nu}(s) \rangle$.
At around the second crossing point, $s = 1 - s_c$, a transition back to the ground state occurs.
After the second transition,
the state adiabatically evolves in the ground state $| \psi^0_{n,\nu}(s) \rangle$ to be $| w(M-1) \rangle$ eventually at $s=1$.
We thus obtain the solution, the name of the right vertex, in polynomial computational time.

Now we show that our conjecture is correct for this algorithm.
Since the algorithm uses the ground and first excited states, we investigate Conjecture.
We take the set $\mathsf{H} = \{(n,\nu) | n=1,2,...\}$ as the collection of all the instances
and define families of ground and first excited states as
\begin{align}
F_{\mathsf H}(s,\lambda) &= \bigl\{ | \psi^\lambda_{n,\nu}(s) \rangle \bigr\}_{(n,\nu) \in \mathsf H},
\notag
\end{align}
where $\lambda = 0,1$.
Then, we find that $p = 2$ for $F_{\mathsf H}(s,0)$
with $s_0 \le s \le s_c$ and $1 - s_c \le s \le 1 - s_0$.
We also find that $p = 2$ for $F_{\mathsf H}(s,1)$
with $s_c \le s < \alpha$ and $1 - \alpha < s \le 1 - s_c$.
Here, $s_0$ is some $O(1)$ constant.
In the following, we show this result.

To this end, we note that it is sufficient to analyze the region of $s_0 \le s < \alpha$
since the Hamiltonian (\ref{Hs_gluedTrees}) has $(s, j) \leftrightarrow (1-s, 2n+1-j)$ symmetry.
As shown in Appendix~\ref{appendix:derivation_psi_gluedTrees},
we can write the ground state for $s_0 \le s \le s_c$
and the first excited state for $s_c \le s < \alpha$ in the following form:
\begin{align}
| \psi^\lambda_{n,\nu}(s) \rangle = c | w(0) \rangle + c_\perp | w(0)^\perp \rangle,
\label{psi_gluedTrees}
\end{align}
where $c$ and $c_\perp = \sqrt{1 - c^2}$ are $O(1)$ real numbers, and
\begin{align}
| w(0)^\perp \rangle = \sum_{v \neq 0} d_v | w(v) \rangle
\label{w_non0}
\end{align}
is a state orthogonal to $| w(0) \rangle$
(the concrete forms of $c$, $c_\perp$, and $d_v$'s are not important here).

To show that $F_{\mathsf H}(s,\lambda)$ has $p=2$,
it is sufficient to find one additive observable whose fluctuation is macroscopically large.
We show that $\hat{A} \equiv \sum_{l=1}^{2n} (-1)^{w_l(0)} \hat{\sigma}_z(l)$
is such an additive operator by using the two facts:
\begin{align}
&\hat{A} | w(0) \rangle = 2n | w(0) \rangle,
\label{A_0_gluedTrees}
\\
&\big| 2n - \langle w(0)^\perp | \hat{A} | w(0)^\perp \rangle \big| = \overline{\Theta}(n).
\label{A_non0_gluedTrees}
\end{align}
Equation~(\ref{A_0_gluedTrees}) is obvious from the definition of $\hat{A}$,
and Eq.~(\ref{A_non0_gluedTrees}) is shown in Appendix~\ref{appendix:derivation_A_non0_gluedTrees}.
These two facts lead to
\begin{align}
&\langle \psi^\lambda_{n,\nu}(s) | \hat{A}^2 | \psi^\lambda_{n,\nu}(s) \rangle
- \langle \psi^\lambda_{n,\nu}(s) | \hat{A} | \psi^\lambda_{n,\nu}(s) \rangle^2
\notag\\
&= c^2 c_\perp^2
\Bigl\{ 2n - \langle w(0)^\perp | \hat{A} | w(0)^\perp \rangle \Bigr\}^2
\notag\\
&~~~+ c_\perp^2 \Bigl\{ \langle w(0)^\perp | \hat{A}^2 | w(0)^\perp \rangle
- \langle w(0)^\perp | \hat{A} | w(0)^\perp \rangle^2 \Bigr\}
\notag\\
&= \overline{\Theta}(n^2).
\end{align}
In the first equality
we used Eq.~(\ref{A_0_gluedTrees}) together with $c_\perp^2 = 1 - c^2$
and $\langle w(0) | w(0)^\perp \rangle = 0$,
and in the last line we used Eq.~(\ref{A_non0_gluedTrees})
with the fact that $c$ and $c_\perp$ are $O(1)$.
Therefore, $\hat{A}$ is a macroscopically-fluctuating additive operator
for $F_{\mathsf H}(s,0)$ with $s_0 \le s \le s_c$
and for $F_{\mathsf H}(s,1)$ with $s_c \le s < \alpha$.
Therefore, we conclude that our conjecture is correct for the quantum adiabatic algorithm for the glued-trees problem.

We may understand this result as follows.
Initially, the state is $| w(0) \rangle$ localized at the left root.
And, as $s$ increases, it gradually becomes delocalized
to be a superposition of the state localized at the left root and the states localized at other vertices [Eq.~(\ref{psi_gluedTrees}) with Eq.~(\ref{w_non0})].
Equations (\ref{A_0_gluedTrees}) and (\ref{A_non0_gluedTrees}) indicate that
the state in Eq.~(\ref{psi_gluedTrees}) is a superposition of macroscopically distinct states for almost every instance.

\section{Concluding remarks}\label{sec:conclusion}

In summary, we have proposed a conjecture on a necessary condition for the quantum speedup in AQC:
if AQC outperforms classical computation in the sense of the strong quantum speedup \cite{Papageorgiou_etal:2013},
then a superposition of macroscopically distinct states (signaled by $p = 2$)
appears during the computation.
We have calculated the index $p$ for the states in five algorithms to test our conjecture.
We have shown that all the results are consistent with our conjecture.
These results suggest that the index $p$ is an appropriate indicator of entanglement
that plays an essential role in the quantum speedup.

As mentioned in Sec.~\ref{sec:index_p},
the index $p$ represents an aspect of entanglement that is not captured by bipartite entanglement measures.
This may lead to difference in the relations between entanglement and quantum speedup.
In fact, in the adiabatic Grover algorithm, the index $p$ can detect the quantum speedup (Sec.~\ref{sec:Grover})
whereas the entanglement entropy cannot detect it \cite{Orus_etal:2004, Latorre_etal:2004}.
As shown in Ref.~\cite{Vidal:2003},
the entanglement entropy is a lower bound on an entanglement measure $E_\chi$ based on the Schmidt rank,
and $E_\chi = O(n^a)$ (with a positive constant $a$) is necessary for the exponential quantum speedup.
This suggests that the entanglement entropy can detect the exponential speedup
but it cannot detect the polynomial speedup.
The index $p$, by contrast, can detect both the exponential and polynomial quantum speedups
(if our conjecture is correct).
Therefore, we expect that the index $p$ is more suitable for investigating the strong quantum speedup.

From the results in Sec.~\ref{sec:examples},
we may classify AQC algorithms (with the quantum speedup) into two.
In the first class, including the adiabatic algorithms for the Grover and glued-trees problems,
there exist crossing points at some $s$ and states with $p=2$ appear at intermediate $s$ ($0<s<1$).
In the second class, including the adiabatic Bernstein--Vazirani and Simon algorithms,
crossing points are absent, and states with $p=2$ appear at the ends of algorithms.
Related to this classification, we can show the following proposition
straightforwardly from the results in Ref.~\cite{Kuwahara_etal:2017}:
``Let $\hat{H}$ be a Hamiltonian satisfying
\begin{align}
\hat{H} = \sum_{X :|X| \le k} \hat{h}_X ~~\mathrm{with}~
\sum_{X : X \ni l} |\hat{h}_X| \le g, ~ \forall l,
\label{localHamiltonian}
\end{align}
where $X$ is a subset of the sites ($|X|$ is the number of the sites included),
and $k$ and $g$ are $O(1)$ constants.
If the ground state of $\hat{H}$ has $p=2$,
then either the energy gap closes in the infinite-size limit or the ground state is degenerate.''
Strictly speaking, we cannot apply this proposition to the examples in Sec.~\ref{sec:examples},
since their Hamiltonians do not satisfy the conditions of Eq.~(\ref{localHamiltonian}).
Nevertheless, we see a rough correspondence between the first (second) class and the former (latter) of the proposition.
Restricted to the first class with Hamiltonians of Eq.~(\ref{localHamiltonian}),
our conjecture (if it is correct) justifies an expectation
that (either power-law or exponential) gap closing, implying quantum phase transition,
is necessary for the quantum speedup in AQC.

Provided that our conjecture is correct for general AQC,
it could make a large contribution to the development of AQC.
For example, in many cases of AQC using stoquastic Hamiltonians,
we do not have conclusive evidence for the strong quantum speedup at present
(see Sec.~VI of Ref.~\cite{Albash_etal:2016} for a review).
In these cases, we may use the index $p$ to judge the speedup.
We note that $p = 2$ does not immediately mean the speedup but $p < 2$ immediately means no speedup.
Of particular interest is to evaluate the index $p$ in AQC for the NP-complete problems,
such as the exact cover problem \cite{Farhi_etal:2001, Young_etal:2010}.
In the exact cover problem, it is difficult to numerically evaluate the energy gap for large-size instances
because the gap becomes smaller than the numerical precision \cite{Young_etal:2010}.
Even in such cases we could evaluate $p$ because $1 \le p \le 2$.

We also discuss the stability of states with $p=2$.
It was shown \cite{Shimizu_etal:2002} that
the decoherence rate of a state with $p$ can be $\Theta(n^p)$ for some noise.
This means that for any state with $p=2$, there is some dangerous noise to which the state is highly unstable.
Therefore, in order to construct a stable AQC,
the protection from the dangerous noise is more important than that from other noises.
Note that we can identify the dangerous noise from the macroscopically-fluctuating additive observable \cite{Shimizu_etal:2002}.

Finally, we remark that we can extend our conjecture to the case of quantum annealing,
which uses mixed states.
For mixed states, we can detect a superposition of macroscopically distinct states by an index $q$,
a natural extension of $p$ \cite{Shimizu_etal:2005, Shimizu_etal:2013}.
Hence we have only to replace $p$ with $q$ in Conjecture.
We should also test the extended conjecture in quantum annealing algorithms (e.g. Ref.~\cite{Amin_etal:2008})
known to achieve the quantum speedup.

\begin{acknowledgments}
The author acknowledges Ryoichi Morita, Satoshi Sasaki, and Daichi Shito
for their help in an early stage of this study.
The author also acknowledges Akira Shimizu and Jun Takahashi for discussions.
This work was supported by JSPS KAKENHI (Grants No. 26287087).
\end{acknowledgments}

\appendix

\section{Derivation of Eq.~(\ref{e_max_BV})}\label{appendix:derivation_of_e_max_BV}

First we consider the case of $a = 0$.
In this case, $f_a(w) = 0$ for any $w \in \{0,1\}^n$,
and hence $| \psi_{n,a}^0(1) \rangle = (1/\sqrt{N}) \sum_{w=0}^{N-1} | w \rangle \otimes | 0 \rangle_{n+1}
= \bigotimes_{l=1}^n | + \rangle_l \otimes | 0 \rangle_{n+1}$.
This is a product state.
Therefore $| \psi_{n,a}^0(1) \rangle$ for $a = 0$ has $e_\mathrm{max} = 2$
since any product states have $e_\mathrm{max} = 2$ \cite{Shimizu_etal:2013}.

Hereafter in this Appendix, we assume $a \ge 1$.
We define $k_a \equiv \sum_{l=1}^n a_l$.
Obviously, $1 \le k_a \le n$.
The goal is to show $e_\mathrm{max}(n,a) = k_a + 1$ for $| \psi_{n,a}^0(1) \rangle$.

\subsection{Another expression of $| \psi_{n,a}^0(1) \rangle$}

We first note that, for fixed $k_a$, we may change the value of $a$ to $a = \sum_{l=n-k_a+1}^n 2^{l-1}$
by simple relabeling of the bits.
Therefore, it is sufficient to analyze the instances of $a = \sum_{l=n-k_a+1}^n 2^{l-1}$ ($k_a = 1, 2, ..., n$).
For such an instance, we can rewrite $| \psi_{n,a}^0(1) \rangle$ in Eq.~(\ref{gs_BV_s1}) as
\begin{align}
| \psi_{n,a}^0(1) \rangle &= \frac{1}{\sqrt{N_a^\prime}} \sum_{w' = 0}^{N_a^\prime - 1} | w' \rangle
\otimes | \varphi_{k_a, \mathrm{e}} \rangle
\notag\\
&= \bigotimes_{l=1}^{n-k_a} | + \rangle_l \otimes | \varphi_{k_a, \mathrm{e}} \rangle,
\label{gs_BV_s1_rewrite_2}
\end{align}
where $N_a^\prime = 2^{n-k_a}$.
Here, the state $| \varphi_{k_a, \mathrm{e}} \rangle$ of the system of qubits for $n-k_a+1 \le l \le n+1$
is recursively defined as
\begin{align}
| \varphi_{k_a, \mathrm{e}} \rangle
&= \frac{1}{\sqrt{2}} \Bigl( | \varphi_{k_a - 1, \mathrm{e}} \rangle \otimes | 0 \rangle
+ | \varphi_{k_a - 1, \mathrm{o}} \rangle \otimes | 1 \rangle \Bigr),
\notag\\
| \varphi_{k_a, \mathrm{o}} \rangle
&= \frac{1}{\sqrt{2}} \Bigl( | \varphi_{k_a - 1, \mathrm{e}} \rangle \otimes | 1 \rangle
+ | \varphi_{k_a - 1, \mathrm{o}} \rangle \otimes | 0 \rangle \Bigr),
\notag
\end{align}
and $| \varphi_{1, \mathrm{e}} \rangle
= \bigl( | 0 \rangle \otimes | 0 \rangle + | 1 \rangle \otimes | 1 \rangle \bigr) / \sqrt{2}$
and $| \varphi_{1, \mathrm{o}} \rangle
= \bigl( | 0 \rangle \otimes | 1 \rangle + | 1 \rangle \otimes | 0 \rangle \bigr) / \sqrt{2}$.
In other words, $| \varphi_{k_a, \mathrm{e}} \rangle$ ($| \varphi_{k_a, \mathrm{o}} \rangle$)
is the uniform superposition of the computational basis states
where even (odd) number of the qubits are in $| 1 \rangle$.

Furthermore, by using the $x$-basis states,
we can show that $| \varphi_{k_a, \mathrm{e}} \rangle$ and $| \varphi_{k_a, \mathrm{o}} \rangle$
are expressed as
\begin{align}
| \varphi_{k_a, \mathrm{e}} \rangle
= \frac{1}{\sqrt{2}} \bigl(  | + + \cdots + \rangle + | - - \cdots - \rangle \bigr),
\label{ka_e}
\\
| \varphi_{k_a, \mathrm{o}} \rangle
= \frac{1}{\sqrt{2}} \bigl(  | + + \cdots + \rangle - | - - \cdots - \rangle \bigr).
\end{align}
That is, $| \varphi_{k_a, \mathrm{e}} \rangle$ is the GHZ state of $k_a + 1$ qubits in the $x$ basis.
Going back to the original labels of the (qu)bits, we obtain Eq.~(\ref{gs_BV_s1_rewrite})
from Eq.~(\ref{gs_BV_s1_rewrite_2}) with Eq.~(\ref{ka_e}).

\subsection{The variance-covariance matrix for $| \psi_{n,a}^0(1) \rangle$}

We next calculate the VCM for $| \psi_{n,a}^0(1) \rangle$.
As seen in Sec.~\ref{sec:VCM}, the matrix elements of the VCM
are the correlations of the Pauli matrices of two qubits.
From Eq.~(\ref{gs_BV_s1_rewrite_2}), it is clear that there are no correlations with the first $n-k_a$ qubits.
Therefore it is sufficient to investigate the correlations among the remaining $k_a + 1$ qubits
in the state $| \varphi_{k_a, \mathrm{e}} \rangle$.

Equation~(\ref{ka_e}) leads to
$\langle \varphi_{k_a, \mathrm{e}} | \hat{\sigma}_\alpha(j) | \varphi_{k_a, \mathrm{e}} \rangle = 0$
($\alpha = x,y,z$).
Therefore, the elements of the VCM can be written as
$V_{j\alpha,j'\alpha'}
= \langle \varphi_{k_a, \mathrm{e}} | \hat{\sigma}_\alpha(j) \hat{\sigma}_{\alpha'}(j') | \varphi_{k_a, \mathrm{e}} \rangle$.
From Eq.~(\ref{ka_e}), these elements can be calculated as
\begin{align}
\langle \varphi_{k_a, \mathrm{e}} | \hat{\sigma}_\alpha(j) \hat{\sigma}_{\alpha'}(j) | \varphi_{k_a, \mathrm{e}} \rangle
&= \delta_{\alpha, \alpha'},
\notag
\end{align}
and for $j \neq j'$
\begin{align}
\langle \varphi_{k_a, \mathrm{e}} | \hat{\sigma}_\alpha(j) \hat{\sigma}_{\alpha'}(j') | \varphi_{k_a, \mathrm{e}} \rangle
&=
\begin{cases}
1 & \alpha = \alpha' = x
\\
0 & \mathrm{otherwise}.
\end{cases}
\notag
\end{align}
Therefore, we may write the VCM, which is a $3(k_a+1) \times 3(k_a+1)$ matrix, as
\begin{align}
V =
\begin{pmatrix}
I_3 & X_3 & X_3 & \cdots & X_3
\\
X_3 & I_3 & X_3 & \cdots & X_3
\\
X_3 & X_3 & I_3 & \cdots & X_3
\\
\vdots & \vdots & \vdots & \ddots & \vdots
\\
X_3 & X_3 & X_3 & \cdots & I_3
\end{pmatrix}.
\notag
\end{align}
Here, $I_3$ is the $3 \times 3$ identity matrix, and
\begin{align}
X_3 =
\begin{pmatrix}
1 & 0 & 0
\\
0 & 0 & 0
\\
0 & 0 & 0
\end{pmatrix}.
\notag
\end{align}
By exchanging the columns and rows, we can transform $V$ into
\begin{align}
V =
\begin{pmatrix}
J_{k_a+1} & 0
\\
0 & I_{2(k_a + 1)}
\end{pmatrix},
\label{VCM_BV}
\end{align}
where $J_{k_a+1}$ is the $(k_a+1) \times (k_a+1)$ matrix all the elements of which are one,
and $I_{2(k_a + 1)}$ is the $2(k_a+1) \times 2(k_a+1)$ identity matrix.

\subsection{The maximum eigenvalue of the VCM}

We next determine the maximum eigenvalue of $V$.
From Eq.~(\ref{VCM_BV}), we may separately calculate the eigenvalues of $J_{k_a+1}$ and $I_{2(k_a + 1)}$.
Obviously, the eigenvalues of $I_{2(k_a + 1)}$ are one.
On the other hand, those of $J_{k_a+1}$ are $k_a + 1$ (non-degenerate) and zero ($k_a$-fold degenerate),
as we show in the next paragraph
Therefore, we show that the maximum eigenvalue of the VCM is $e_\mathrm{max}(n,a) = k_a + 1$.

Here we determine the eigenvalues of $J_{k_a+1}$
by showing that the characteristic polynomial
$P_{k_a}(\lambda) \equiv \mathrm{det} [J_{k_a+1} - \lambda I_{k_a+1}]$
is written as $P_{k_a}(\lambda) = (-\lambda)^{k_a} (k_a + 1 - \lambda)$.
We can show this as follows.
From the definition of $P_{k_a}(\lambda)$, we have
$P_{k_a}(\lambda) = -\lambda \bigl[2 P_{k_a-1}(\lambda) + \lambda P_{k_a-2}(\lambda) \bigr]$.
This recurrence relation together with $P_0(\lambda) = 1 - \lambda$ and
$P_1(\lambda) = -\lambda (2 - \lambda)$---these can be verified by direct calculation---leads to
that $P_{k_a}(\lambda) = (-\lambda)^{k_a} (k_a + 1 - \lambda)$ is valid for $k_a \ge 0$.
Therefore we show that the eigenvalues of $J_{k_a+1}$ are $k_a + 1$ and zero.

In addition, we can easily show that $\bm{u}^\mathrm{max} \equiv (1, 1, ..., 1)$
[$(k_a+1)$-dimensional unit vector] satisfies the eigenvalue equation:
$J_{k_a+1} \bm{u}^\mathrm{max} = (k_a + 1) \bm{u}^\mathrm{max}$.
From this, we identify the macroscopically-fluctuating observable
as $\hat{A} = \sum_{l=1}^n a_l \hat{\sigma}_x(l) + \hat{\sigma}_x(n+1)$.

\section{Derivation of Eq.~(\ref{e_max_Simon})}\label{appendix:derivation_of_e_max_Simon}

We define $k_a \equiv \sum_{l=1}^n a_l$.
First we consider the case of $a = 2^{l-1}$ ($l = 1,2,...,n$).
In this case, $k_a = 1$ and we can rewrite Eq.~(\ref{state_Simon}) as
\begin{align}
| \psi_{n,a,w^*} \rangle
= \bigotimes_{\scriptstyle l_1 = 1 \atop \scriptstyle (l_1 \neq l(a))}^n | w^*_{l_1} \rangle_{l_1}
\otimes | + \rangle_{l(a)} \otimes | g(w^*) \rangle,
\end{align}
where $l(a)$ is the label satisfying $a_{l(a)} = 1$.
This is a product state and thus has  $e_\mathrm{max} = 2$
since any product states have $e_\mathrm{max} = 2$ \cite{Shimizu_etal:2013}.

Hereafter, we assume $2 \le k_a \le n$.
We first note that, for fixed $k_a$,
we may change the value of $a$ to $a = \sum_{l_1=1}^{k_a} 2^{l_1 - 1}$ by simple relabeling of the bits.
Therefore it is sufficient to analyze the instances of $a = \sum_{l_1=1}^{k_a} 2^{l_1 - 1}$ ($k_a = 2, 3, ..., n$).
For such an instance, we can rewrite Eq.~(\ref{state_Simon}) as
\begin{align}
| \psi_{n,a,w^*} \rangle
&= | \varphi_{k_a,w^*} \rangle \bigotimes_{l_1 = k_a + 1}^n | w^*_{l_1} \rangle_{l_1} \otimes | g(w^*) \rangle,
\label{state_Simon_rewrite}
\\
| \varphi_{k_a,w^*} \rangle
&= \frac{1}{\sqrt{2}} \bigl( | w^{k_a} \rangle + | \tilde{w}^{k_a} \rangle \bigr),
\label{state_Simon_ka}
\end{align}
where $| w^{k_a} \rangle = \bigotimes_{l_1 = 1}^{k_a} | w^*_{l_1} \rangle_{l_1}$
is the first $k_a$-qubit extraction of $| w^* \rangle$,
and $| \tilde{w}^{k_a} \rangle = \bigotimes_{l_1 = 1}^{k_a} | 1 - w^*_{l_1} \rangle_{l_1}$
is the all-bit-flipped state of $| w^{k_a} \rangle$.

We next calculate the VCM for $| \psi_{n,a,w^*} \rangle$.
The matrix elements of the VCM are the correlations of the Pauli matrices of two qubits.
Equation~(\ref{state_Simon_rewrite}) implies that there are no correlations
between the qubits with labels $k_a + 1 \le l_1 \le n$ (in the first subsystem)
and those with $1 \le l_2 \le n - 1$ (in the second subsystem).
Therefore it is sufficient to investigate the correlations among the first $k_a$ qubits
in the state $| \varphi_{k_a,w^*} \rangle$.

Using Eq.~(\ref{state_Simon_ka}),
we can calculate the expectation values of $\hat{\sigma}_\alpha(l_1)$ ($\alpha=x,y,z$) as
$\langle \varphi_{k_a,w^*} | \hat{\sigma}_\alpha(l_1) | \varphi_{k_a,w^*}\rangle = 0$.
Also we have
$\langle \varphi_{k_a,w^*} | \hat{\sigma}_\alpha(l_1) \hat{\sigma}_{\alpha'}(l_1) | \varphi_{k_a,w^*}\rangle
= \delta_{\alpha, \alpha'}$
and
\begin{align}
\langle \varphi_{k_a,w^*} | \hat{\sigma}_\alpha(l_1) \hat{\sigma}_{\alpha'}(l_1') | \varphi_{k_a,w^*}\rangle =
\begin{cases}
(-1)^{w^*_{l_1} + w^*_{l_1'}} & \alpha = \alpha' = z
\\
0 & \mathrm{otherwise},
\end{cases}
\notag
\end{align}
for $l_1 \neq l_1'$.
Hence we may write the VCM, which is a $3k_a \times 3k_a$ matrix, as
\begin{align}
V =
\begin{pmatrix}
I_3 & Z_3^{(1,2)} & Z_3^{(1,3)} & \cdots & Z_3^{(1,k_a)}
\\
Z_3^{(2,1)} & I_3 & Z_3^{(2,3)} & \cdots & Z_3^{(2,k_a)}
\\
Z_3^{(3,1)} & Z_3^{(3,2)} & I_3 & \cdots & Z_3^{(3,k_a)}
\\
\vdots & \vdots & \vdots & \ddots & \vdots
\\
Z_3^{(k_a,1)} & Z_3^{(k_a,2)} & Z_3^{(k_a,3)} & \cdots & I_3
\end{pmatrix}.
\notag
\end{align}
Here, $I_3$ is the $3 \times 3$ identity matrix, and
\begin{align}
Z_3^{(l,l')} =
\begin{pmatrix}
0 & 0 & 0
\\
0 & 0 & 0
\\
0 & 0 & (-1)^{w^*_l + w^*_{l'}}
\end{pmatrix}.
\notag
\end{align}
By exchanging the columns and rows, we can transform $V$ into
\begin{align}
V =
\begin{pmatrix}
J'_{k_a} & 0
\\
0 & I_{2k_a}
\end{pmatrix},
\label{VCM_Simon}
\end{align}
where $J'_{k_a}$ is the $k_a \times k_a$ matrix whose $(l,l')$ element is $(-1)^{w^*_l + w^*_{l'}}$,
and $I_{2k_a}$ is the $2k_a \times 2k_a$ identity matrix.

Therefore, in a similar manner to that in Appendix~\ref{appendix:derivation_of_e_max_BV},
we obtain the maximum eigenvalue of $V$ as $e_\mathrm{max}(n,a) = k_a$.
In addition, we identify the macroscopically-fluctuating observable as
$\hat{A} = \sum_{l_1 = 1}^n (-1)^{w^*_{l_1}} a_{l_1} \hat{\sigma}_z(l_1)$.

\section{Rigorous estimation for asymptotic behavior of binomially distributed $e_\mathrm{max}(n,a)$}
\label{appendix:estimation_of_e_max}

Here we show that $e_\mathrm{max}(n,a)$ given by Eq.~(\ref{e_max_BV}) or (\ref{e_max_Simon})
asymptotically behaves as
\begin{align}
e_\mathrm{max}(n,a) = \overline{\Theta}(n),
\label{e_max_asym_appendix}
\end{align}
for the family of all the instances.

We define $k_a(n) \equiv \sum_{l=1}^n a_l$ for $a \in \{0,1\}^n$.
Then, from Eq.~(\ref{e_max_BV}), we may say that
\begin{align}
e_\mathrm{max}(n,a) = k_a(n) + 1
\label{e_max_BV_appendix}
\end{align}
holds for almost every $a$ in the adiabatic Bernstein--Vazirani algorithm.
And, from Eq.~(\ref{e_max_Simon}), we may say that
\begin{align}
e_\mathrm{max}(n,a) = k_a(n)
\label{e_max_Simon_appendix}
\end{align}
holds for almost every $a$ in the adiabatic Simon algorithm.
The number of instances $N_\mathrm{not}$ \textit{not} satisfying Eq.~(\ref{e_max_BV_appendix}) or (\ref{e_max_Simon_appendix}) is vanishingly small:
$\lim_{n \to \infty} N_\mathrm{not} / N = 0$,
where $N = 2^n$ is the total number of $a$'s (instances) for fixed $n$.

Therefore, in the following, we consider the instances that satisfy
Eq.~(\ref{e_max_BV_appendix}) or (\ref{e_max_Simon_appendix}).
For such instances, we verify Eq.~(\ref{e_max_asym_appendix}) by showing that,
for any $\varepsilon$ ($0< \varepsilon < 1$),
$(1 - \varepsilon)n/2 \le k_a(n) \le (1 + \varepsilon)n/2$ holds
for almost every $a \in \{0,1\}^n$ (for sufficiently large $n$).

In the above statement, ``almost every $a$'' means
\begin{align}
\lim_{n \to \infty} \frac{N_\mathrm{exc}(n,\varepsilon)}{N} = 0,
\label{exceptions_vanish}
\end{align}
where $N_\mathrm{exc}(n,\varepsilon)$ is the number of the exceptional instances
that satisfy $\big| k_a(n) - n/2 \big| > \varepsilon n/2$.
Therefore the goal of this Appendix is to show Eq.~(\ref{exceptions_vanish}).

We can show Eq.~(\ref{exceptions_vanish}) as follows.
The number of the instances that satisfy $k_a(n) = k$ is equal to
the binomial coefficient $\binom{n}{k} = n! / [k! (n-k)!]$.
Thanks to the Stirling formula \cite{WangGuo}, we can show that for $m \ge 1$
\begin{align}
\sqrt{2\pi m} (m/e)^m < m! < \sqrt{2\pi m} (m/e)^m e^{1/12m}.
\notag
\end{align}
Therefore we may estimate $\binom{n}{k}$ for $1 \le k \le n-1$ as
\begin{align}
\binom{n}{k} < \frac{n^{n+1/2} e^{1/12n}}{\sqrt{2\pi} k^{k+1/2} (n-k)^{n-k+1/2}}.
\notag
\end{align}
Furthermore, if $k = \lceil (1 - \varepsilon)n/2 \rceil$ or $k = \lfloor (1 + \varepsilon)n/2 \rfloor$
($0 < \varepsilon < 1$), we may estimate the above equation as
\begin{align}
\binom{n}{k} &< \sqrt{\frac{2}{\pi (1 - \varepsilon^2) n}} N
\exp \left[ -\frac{1}{2} h(\varepsilon) n + \frac{1}{12n} \right],
\notag
\end{align}
where $h(\varepsilon) = (1 + \varepsilon) \log (1 + \varepsilon) + (1 - \varepsilon) \log (1 - \varepsilon)$.

Using the above equation, we can estimate $N_\mathrm{exc}$ as
\begin{align}
&N_\mathrm{exc}(n,\varepsilon) = \sum_{k = 0}^{\lfloor (1 - \varepsilon)n/2 \rfloor} \binom{n}{k} +
\sum_{k = \lceil (1 + \varepsilon)n/2 \rceil}^{n} \binom{n}{k}
\notag
\\
&< 2 + (1 - \varepsilon)n \sqrt{\frac{2}{\pi (1 - \varepsilon^2) n}} N
\exp \left[ -\frac{1}{2} h(\varepsilon) n + \frac{1}{12n} \right],
\notag
\end{align}
where the first term in the last line follows from $\binom{n}{0} = \binom{n}{n} = 1$.
Dividing this equation by $N$, we finally obtain
\begin{align}
\frac{N_\mathrm{exc}(n,\varepsilon)}{N} < \frac{2}{N} + \sqrt{\frac{2(1 - \varepsilon) n}{\pi (1 + \varepsilon)}}
\exp \left[ -\frac{1}{2} h(\varepsilon) n + \frac{1}{12n} \right].
\end{align}
The right hand side approaches zero as $n \to \infty$ because $h(\varepsilon) > 0$ for $0 < \varepsilon < 1$.

\section{Derivation of Eq.~(\ref{psi_gluedTrees})}
\label{appendix:derivation_psi_gluedTrees}

Here, we derive the ground and first excited states of the Hamiltonian (\ref{Hs_gluedTrees})
in the glued-trees problem.
This appendix is based on the supplemental material of Ref.~\cite{Somma_etal:2012}.
New results here are explicit (approximate) forms of the states,
Eqs.~(\ref{psi_gluedTrees_s}) and (\ref{psi_gluedTrees_sc})
(as well as the energy gap at $s_c$).

In this appendix, we assume $0 < s < \alpha$.
For convenience, we rescale the Hamiltonian as
$\hat{H}' = \hat{H}(s;n,\nu) / [s(1-s)]
= \alpha' \hat{H}_\mathrm{ini} + \beta' \hat{H}_\mathrm{fin} - \hat{H}_\mathrm{ora}$,
where $\alpha' = \alpha / s$ and $\beta' = \alpha / (1-s)$.
Following the supplemental material of Ref.~\cite{Somma_etal:2012},
we can write the ground and first excited states in forms of
$| \psi \rangle = \sum_{j=0}^{2n+1} \gamma_j | \mathrm{col}(j) \rangle$, where
\begin{align}
\gamma_j =
\begin{cases}
a e^{qj} + b e^{-qj} & 0 \le j \le n
\\
c e^{q(2n+1 - j)} + d e^{-q(2n+1 - j)} & n+1 \le j \le 2n+1
\end{cases}
\label{coefficient_ansatz}
\end{align}
with real constants $a, b, c, d, q$ to be determined ($q > 0$).
From the eigenvalue equation, $\hat{H}' | \psi \rangle = E' | \psi \rangle$,
we obtain $E' = - 2 \mathrm{cosh} q$ and
\begin{align}
\mathcal{M} & (a,b,c,d)^T = 0,
\label{linearEq_AppendixD}
\\
\mathcal{M} &=
\begin{pmatrix}
e^{-q} - \alpha' & e^q - \alpha' & 0 & 0
\\
0 & 0 & e^{-q} - \beta' & e^q - \beta'
\\
e^{q(n+1)} & e^{-q(n+1)} & -\sqrt{2} e^{qn} & -\sqrt{2} e^{-qn}
\\
-\sqrt{2} e^{qn} & -\sqrt{2} e^{-qn} & e^{q(n+1)} & e^{-q(n+1)}
\end{pmatrix}.
\notag
\end{align}
Since Eq.~(\ref{linearEq_AppendixD}) has non-trivial solutions,
$\mathrm{det} \mathcal{M} = 0$, which leads to
\begin{align}
&(1 - \alpha' x) (x - \beta') x^{-2n+3}
+ (x - \alpha') (1 - \beta' x) x^{-2n+3}
\notag\\
&+ (1 + \sqrt{2} x) (1 - \sqrt{2} x) (1 - \alpha' x) (1 - \beta' x) x^{-4n}
\notag\\
&+ (x + \sqrt{2}) (x - \sqrt{2}) (x - \alpha') (x - \beta') x^4
= 0,
\label{detM=0}
\end{align}
with $x = e^q$.
The solutions of Eq.~(\ref{detM=0}) in the limit of $n \to \infty$
are $x_\infty = \sqrt{2}$ and $x_\infty = \alpha'$
[note that $q > 0$ ($x > 1$) and $\beta' < 1$].
The state corresponding to $x_\infty = \alpha'$ is the ground state for $s < s_c$
and is the first excited state for $s_c < s~(< \alpha)$
because $E' = - 2 \mathrm{cosh}q = -(x + x^{-1})$.
At $s=s_c$ ($\alpha' = \sqrt{2}$), in particular, these two solutions are equal to each other,
which indicates the level crossing in the infinite-size limit.
In the below, we analyze the finite-size correction for $s \neq s_c$ and $s = s_c$ separately.

First, we consider the case of $s \neq s_c$.
In this case, it is reasonable to expand the solution of Eq.~(\ref{detM=0})
in powers of $\epsilon' \equiv (\alpha')^{-n}$.
We substitute $x = \alpha' + x_1 \epsilon' + O(\epsilon^{\prime 2})$ into Eq.~(\ref{detM=0})
and collect the terms proportional to $\epsilon'$ to obtain $x_1 = 0$.

We may determine the corresponding eigenstates
by solving Eq.~(\ref{linearEq_AppendixD}) in powers of $\epsilon'$
and by using the normalization condition with Eq.~(\ref{coefficient_ansatz}).
From the condition $|\gamma_j| \le 1$ for Eq.~(\ref{coefficient_ansatz}),
the leading terms of $(a, b, c, d)$ should be $(a_1 \epsilon', b_0, c_1 \epsilon', d_0)$.
We substitute these leading terms into Eq.~(\ref{linearEq_AppendixD})
with first-order approximation of $\mathcal{M}$ in $\epsilon'$.
Then, from the terms proportional to $\epsilon^{\prime 0}$ and $\epsilon'$,
we can show $a_1 = c_1 = d_0 = 0$.
From this result, we obtain the eigenstate corresponding to $x = \alpha' + O(\epsilon^{\prime 2})$
within an error of $O(\epsilon^{\prime 2})$:
\begin{align}
| \psi \rangle &\simeq b_0 \sum_{j=0}^n (\alpha/s)^{-j} | \mathrm{col}(j) \rangle
\label{psi_gluedTrees_s}
\\
&= b_0 \sum_{j=0}^n \sum_{v \in V_j} (\sqrt{2}\alpha/s)^{-j} | w(v) \rangle.
\notag
\end{align}
If $s = O(1)$, we can write this result in the form of Eq.~(\ref{psi_gluedTrees})
with $O(1)$ constants $c$ and $c_\perp$.

Next, we consider the case of $s = s_c$.
In this case, it is reasonable to expand the solution of Eq.~(\ref{detM=0})
in powers of $\epsilon \equiv 2^{-n/2}$.
We substitute $x = \sqrt{2} + x_1 \epsilon + O(\epsilon^2)$ into Eq.~(\ref{detM=0})
and collect the terms proportional to $\epsilon^2$ to obtain $x_1 = \pm 1/2$.
Here, the positive (negative) sign corresponds to the ground (first excited) state because $E' = -(x + x^{-1})$.
Therefore, the energy gap of the non-rescaled Hamiltonian at $s_c$
is $\Delta E = s_c (1 - s_c) \epsilon / 2 + O(\epsilon^2)$.

We may determine the corresponding eigenstates in a similar manner to that in the case of $s \neq s_c$.
The result is $d_0 = 0$ and $a_1 = c_1 = \pm b_0/\sqrt{2}$ (corresponding to $x_1 = \pm 1/2$)
for the leading terms of $(a, b, c, d) = (a_1 \epsilon, b_0, c_1 \epsilon, d_0)$.
From this result, we can determine explicit forms of the ground and first excited states within errors of $O(\epsilon^2)$:
\begin{align}
| \psi_\pm \rangle \simeq b_0 \left( \sum_{j=0}^n 2^{-j/2} | \mathrm{col}(j) \rangle
\pm \epsilon \sum_{j=0}^{2n+1} \sqrt{\frac{N_j}{2}} | \mathrm{col}(j) \rangle \right).
\label{psi_gluedTrees_sc}
\end{align}
Here, $N_j = 2^j$ for $0 \le j \le n$ and $N_j = 2^{2n+1 - j}$ for $n+1 \le j \le 2n+1$.
We can write Eq.~(\ref{psi_gluedTrees_sc}) in the form of Eq.~(\ref{psi_gluedTrees}) with $O(1)$ constants $c$ and $c_\perp$.

\section{Derivation of Eq.~(\ref{A_non0_gluedTrees})}
\label{appendix:derivation_A_non0_gluedTrees}

In the expression (\ref{w_non0}) of $| w(0)^\perp \rangle$,
$W_\perp = \{ w(v) \}_{v \neq 0}$ is a set of $M-1$ strings
randomly chosen without repetition from all the $2n$-bit strings except for $w(0)$,
where $M=2^{n+2} - 2$.
Let $P$ be the probability that $W_\perp$ contains
at least one $w(v)$ satisfying $|2n - A_{w(v)}| = o(n)$.
Here, $A_{w(v)}$ is given by $\hat{A} | w(v) \rangle = A_{w(v)} | w(v) \rangle$
with $\hat{A} = \sum_{l=1}^{2n} (-1)^{w_l(0)} \hat{\sigma}_z(l)$.
To derive Eq.~(\ref{A_non0_gluedTrees}),
it is sufficient to show that $P$ is vanishingly small for large $n$.

The number of strings that satisfy $|2n - A_w| = o(n)$
is $L = \sum_{k=1}^K \binom{2n}{k}$ with $K = o(n)$.
This is much smaller than $N = 2^{2n}$.
In practice, we can show, in a similar manner to that in Appendix~\ref{appendix:estimation_of_e_max},
that
\begin{align}
\frac{L M}{N} \to 0 ~ \mathrm{and} ~ \frac{L^2 M}{N} \to 0 ~ \mathrm{as} ~ n \to \infty.
\label{L_is_small}
\end{align}

If $W_\perp$ contains $\mu$ strings satisfying $|2n - A_{w(v)}| = o(n)$,
then the $\mu$ strings are chosen from the above $L$ candidates
and the others are chosen from the rest.
Therefore the probability $P$ is given by
\begin{align}
P = \frac{\sum_{\mu=1}^{L} \binom{L}{\mu} \binom{N-L-1}{M-\mu-1}}{\binom{N-1}{M-1}}.
\notag
\end{align}
We may evaluate this as
\begin{align}
P < \frac{L \binom{L}{\bar{\mu}} \binom{N-L-1}{M-\bar{\mu}-1}}{\binom{N-1}{M-1}}
< \frac{L^{\bar{\mu} + 1} M^{\bar{\mu}}}{(N-L)^{\bar{\mu}}},
\label{P_small}
\end{align}
where $\bar{\mu} = \mathrm{argmax}_{1 \le \mu \le L} \binom{L}{\mu} \binom{N-L-1}{M-\mu-1}$.
The right hand side of Eq.~(\ref{P_small}) approaches zero as $n \to \infty $
due to Eq.~(\ref{L_is_small}).


\begin{thebibliography}{81}%
\makeatletter
\providecommand \@ifxundefined [1]{%
 \@ifx{#1\undefined}
}%
\providecommand \@ifnum [1]{%
 \ifnum #1\expandafter \@firstoftwo
 \else \expandafter \@secondoftwo
 \fi
}%
\providecommand \@ifx [1]{%
 \ifx #1\expandafter \@firstoftwo
 \else \expandafter \@secondoftwo
 \fi
}%
\providecommand \natexlab [1]{#1}%
\providecommand \enquote  [1]{``#1''}%
\providecommand \bibnamefont  [1]{#1}%
\providecommand \bibfnamefont [1]{#1}%
\providecommand \citenamefont [1]{#1}%
\providecommand \href@noop [0]{\@secondoftwo}%
\providecommand \href [0]{\begingroup \@sanitize@url \@href}%
\providecommand \@href[1]{\@@startlink{#1}\@@href}%
\providecommand \@@href[1]{\endgroup#1\@@endlink}%
\providecommand \@sanitize@url [0]{\catcode `\\12\catcode `\$12\catcode
  `\&12\catcode `\#12\catcode `\^12\catcode `\_12\catcode `\%12\relax}%
\providecommand \@@startlink[1]{}%
\providecommand \@@endlink[0]{}%
\providecommand \url  [0]{\begingroup\@sanitize@url \@url }%
\providecommand \@url [1]{\endgroup\@href {#1}{\urlprefix }}%
\providecommand \urlprefix  [0]{URL }%
\providecommand \Eprint [0]{\href }%
\providecommand \doibase [0]{http://dx.doi.org/}%
\providecommand \selectlanguage [0]{\@gobble}%
\providecommand \bibinfo  [0]{\@secondoftwo}%
\providecommand \bibfield  [0]{\@secondoftwo}%
\providecommand \translation [1]{[#1]}%
\providecommand \BibitemOpen [0]{}%
\providecommand \bibitemStop [0]{}%
\providecommand \bibitemNoStop [0]{.\EOS\space}%
\providecommand \EOS [0]{\spacefactor3000\relax}%
\providecommand \BibitemShut  [1]{\csname bibitem#1\endcsname}%
\let\auto@bib@innerbib\@empty
\bibitem [{\citenamefont {Farhi}\ \emph {et~al.}(2000)\citenamefont {Farhi},
  \citenamefont {Goldstone}, \citenamefont {Gutmann},\ and\ \citenamefont
  {Sipser}}]{Farhi_etal:2000}%
  \BibitemOpen
  \bibfield  {author} {\bibinfo {author} {\bibfnamefont {E.}~\bibnamefont
  {Farhi}}, \bibinfo {author} {\bibfnamefont {J.}~\bibnamefont {Goldstone}},
  \bibinfo {author} {\bibfnamefont {S.}~\bibnamefont {Gutmann}}, \ and\
  \bibinfo {author} {\bibfnamefont {M.}~\bibnamefont {Sipser}},\ }\href
  {https://arxiv.org/abs/quant-ph/0001106} {\enquote {\bibinfo {title} {Quantum
  computation by adiabatic evolution},}\ } (\bibinfo {year} {2000}),\ \Eprint
  {http://arxiv.org/abs/quant-ph/0001106} {arXiv:quant-ph/0001106} \BibitemShut
  {NoStop}%
\bibitem [{\citenamefont {Farhi}\ \emph {et~al.}(2001)\citenamefont {Farhi},
  \citenamefont {Goldstone}, \citenamefont {Gutmann}, \citenamefont {Lapan},
  \citenamefont {Lundgren},\ and\ \citenamefont {Preda}}]{Farhi_etal:2001}%
  \BibitemOpen
  \bibfield  {author} {\bibinfo {author} {\bibfnamefont {E.}~\bibnamefont
  {Farhi}}, \bibinfo {author} {\bibfnamefont {J.}~\bibnamefont {Goldstone}},
  \bibinfo {author} {\bibfnamefont {S.}~\bibnamefont {Gutmann}}, \bibinfo
  {author} {\bibfnamefont {J.}~\bibnamefont {Lapan}}, \bibinfo {author}
  {\bibfnamefont {A.}~\bibnamefont {Lundgren}}, \ and\ \bibinfo {author}
  {\bibfnamefont {D.}~\bibnamefont {Preda}},\ }\href {\doibase
  10.1126/science.1057726} {\bibfield  {journal} {\bibinfo  {journal}
  {Science}\ }\textbf {\bibinfo {volume} {292}},\ \bibinfo {pages} {472}
  (\bibinfo {year} {2001})}\BibitemShut {NoStop}%
\bibitem [{\citenamefont {Apolloni}\ \emph {et~al.}(1989)\citenamefont
  {Apolloni}, \citenamefont {Carvalho},\ and\ \citenamefont
  {de~Falco}}]{Apolloni_etal:1989}%
  \BibitemOpen
  \bibfield  {author} {\bibinfo {author} {\bibfnamefont {B.}~\bibnamefont
  {Apolloni}}, \bibinfo {author} {\bibfnamefont {C.}~\bibnamefont {Carvalho}},
  \ and\ \bibinfo {author} {\bibfnamefont {D.}~\bibnamefont {de~Falco}},\
  }\href {\doibase http://dx.doi.org/10.1016/0304-4149(89)90040-9} {\bibfield
  {journal} {\bibinfo  {journal} {Stochastic Processes and their Applications}\
  }\textbf {\bibinfo {volume} {33}},\ \bibinfo {pages} {233 } (\bibinfo {year}
  {1989})}\BibitemShut {NoStop}%
\bibitem [{\citenamefont {Somorjai}(1991)}]{Somorjai:1991}%
  \BibitemOpen
  \bibfield  {author} {\bibinfo {author} {\bibfnamefont {R.~L.}\ \bibnamefont
  {Somorjai}},\ }\href {\doibase 10.1021/j100163a045} {\bibfield  {journal}
  {\bibinfo  {journal} {J. Phys. Chem.}\ }\textbf {\bibinfo {volume} {95}},\
  \bibinfo {pages} {4141} (\bibinfo {year} {1991})}\BibitemShut {NoStop}%
\bibitem [{\citenamefont {Amara}\ \emph {et~al.}(1993)\citenamefont {Amara},
  \citenamefont {Hsu},\ and\ \citenamefont {Straub}}]{Amara_etal:1993}%
  \BibitemOpen
  \bibfield  {author} {\bibinfo {author} {\bibfnamefont {P.}~\bibnamefont
  {Amara}}, \bibinfo {author} {\bibfnamefont {D.}~\bibnamefont {Hsu}}, \ and\
  \bibinfo {author} {\bibfnamefont {J.~E.}\ \bibnamefont {Straub}},\ }\href
  {\doibase 10.1021/j100127a023} {\bibfield  {journal} {\bibinfo  {journal} {J.
  Phys. Chem.}\ }\textbf {\bibinfo {volume} {97}},\ \bibinfo {pages} {6715}
  (\bibinfo {year} {1993})}\BibitemShut {NoStop}%
\bibitem [{\citenamefont {Finnila}\ \emph {et~al.}(1994)\citenamefont
  {Finnila}, \citenamefont {Gomez}, \citenamefont {Sebenik}, \citenamefont
  {Stenson},\ and\ \citenamefont {Doll}}]{Finnila_etal:1994}%
  \BibitemOpen
  \bibfield  {author} {\bibinfo {author} {\bibfnamefont {A.}~\bibnamefont
  {Finnila}}, \bibinfo {author} {\bibfnamefont {M.}~\bibnamefont {Gomez}},
  \bibinfo {author} {\bibfnamefont {C.}~\bibnamefont {Sebenik}}, \bibinfo
  {author} {\bibfnamefont {C.}~\bibnamefont {Stenson}}, \ and\ \bibinfo
  {author} {\bibfnamefont {J.}~\bibnamefont {Doll}},\ }\href {\doibase
  http://dx.doi.org/10.1016/0009-2614(94)00117-0} {\bibfield  {journal}
  {\bibinfo  {journal} {Chem. Phys. Lett.}\ }\textbf {\bibinfo {volume}
  {219}},\ \bibinfo {pages} {343 } (\bibinfo {year} {1994})}\BibitemShut
  {NoStop}%
\bibitem [{\citenamefont {Kadowaki}\ and\ \citenamefont
  {Nishimori}(1998)}]{Kadowaki_etal:1998}%
  \BibitemOpen
  \bibfield  {author} {\bibinfo {author} {\bibfnamefont {T.}~\bibnamefont
  {Kadowaki}}\ and\ \bibinfo {author} {\bibfnamefont {H.}~\bibnamefont
  {Nishimori}},\ }\href {\doibase 10.1103/PhysRevE.58.5355} {\bibfield
  {journal} {\bibinfo  {journal} {Phys. Rev. E}\ }\textbf {\bibinfo {volume}
  {58}},\ \bibinfo {pages} {5355} (\bibinfo {year} {1998})}\BibitemShut
  {NoStop}%
\bibitem [{\citenamefont {Santoro}\ and\ \citenamefont
  {Tosatti}(2006)}]{Santoro_etal:2006}%
  \BibitemOpen
  \bibfield  {author} {\bibinfo {author} {\bibfnamefont {G.~E.}\ \bibnamefont
  {Santoro}}\ and\ \bibinfo {author} {\bibfnamefont {E.}~\bibnamefont
  {Tosatti}},\ }\href {http://stacks.iop.org/0305-4470/39/i=36/a=R01}
  {\bibfield  {journal} {\bibinfo  {journal} {J. Phys. A}\ }\textbf {\bibinfo
  {volume} {39}},\ \bibinfo {pages} {R393} (\bibinfo {year}
  {2006})}\BibitemShut {NoStop}%
\bibitem [{\citenamefont {Das}\ and\ \citenamefont
  {Chakrabarti}(2008)}]{Das_etal:2008}%
  \BibitemOpen
  \bibfield  {author} {\bibinfo {author} {\bibfnamefont {A.}~\bibnamefont
  {Das}}\ and\ \bibinfo {author} {\bibfnamefont {B.~K.}\ \bibnamefont
  {Chakrabarti}},\ }\href {\doibase 10.1103/RevModPhys.80.1061} {\bibfield
  {journal} {\bibinfo  {journal} {Rev. Mod. Phys.}\ }\textbf {\bibinfo {volume}
  {80}},\ \bibinfo {pages} {1061} (\bibinfo {year} {2008})}\BibitemShut
  {NoStop}%
\bibitem [{\citenamefont {Morita}\ and\ \citenamefont
  {Nishimori}(2008)}]{Morita_etal:2008}%
  \BibitemOpen
  \bibfield  {author} {\bibinfo {author} {\bibfnamefont {S.}~\bibnamefont
  {Morita}}\ and\ \bibinfo {author} {\bibfnamefont {H.}~\bibnamefont
  {Nishimori}},\ }\href {\doibase 10.1063/1.2995837} {\bibfield  {journal}
  {\bibinfo  {journal} {J. Math. Phys.}\ }\textbf {\bibinfo {volume} {49}},\
  \bibinfo {pages} {125210} (\bibinfo {year} {2008})}\BibitemShut {NoStop}%
\bibitem [{\citenamefont {Bapst}\ \emph {et~al.}(2013)\citenamefont {Bapst},
  \citenamefont {Foini}, \citenamefont {Krzakala}, \citenamefont {Semerjian},\
  and\ \citenamefont {Zamponi}}]{Bapst_etal:2013}%
  \BibitemOpen
  \bibfield  {author} {\bibinfo {author} {\bibfnamefont {V.}~\bibnamefont
  {Bapst}}, \bibinfo {author} {\bibfnamefont {L.}~\bibnamefont {Foini}},
  \bibinfo {author} {\bibfnamefont {F.}~\bibnamefont {Krzakala}}, \bibinfo
  {author} {\bibfnamefont {G.}~\bibnamefont {Semerjian}}, \ and\ \bibinfo
  {author} {\bibfnamefont {F.}~\bibnamefont {Zamponi}},\ }\href {\doibase
  https://doi.org/10.1016/j.physrep.2012.10.002} {\bibfield  {journal}
  {\bibinfo  {journal} {Physics Reports}\ }\textbf {\bibinfo {volume} {523}},\
  \bibinfo {pages} {127 } (\bibinfo {year} {2013})}\BibitemShut {NoStop}%
\bibitem [{\citenamefont {Suzuki}\ and\ \citenamefont
  {Das}(2015)}]{Suzuki_etal:2015}%
  \BibitemOpen
  \bibfield  {author} {\bibinfo {author} {\bibfnamefont {S.}~\bibnamefont
  {Suzuki}}\ and\ \bibinfo {author} {\bibfnamefont {A.}~\bibnamefont {Das}},\
  }\href {\doibase 10.1140/epjst/e2015-02336-2} {\bibfield  {journal} {\bibinfo
   {journal} {The European Physical Journal Special Topics}\ }\textbf {\bibinfo
  {volume} {224}},\ \bibinfo {pages} {1} (\bibinfo {year} {2015})}\BibitemShut
  {NoStop}%
\bibitem [{\citenamefont {Albash}\ and\ \citenamefont
  {Lidar}(2016)}]{Albash_etal:2016}%
  \BibitemOpen
  \bibfield  {author} {\bibinfo {author} {\bibfnamefont {T.}~\bibnamefont
  {Albash}}\ and\ \bibinfo {author} {\bibfnamefont {D.~A.}\ \bibnamefont
  {Lidar}},\ }\href {https://arxiv.org/abs/1611.04471} {\enquote {\bibinfo
  {title} {Adiabatic quantum computing},}\ } (\bibinfo {year} {2016}),\ \Eprint
  {http://arxiv.org/abs/1611.04471} {arXiv:1611.04471} \BibitemShut {NoStop}%
\bibitem [{\citenamefont {Kato}(1950)}]{Kato:1950}%
  \BibitemOpen
  \bibfield  {author} {\bibinfo {author} {\bibfnamefont {T.}~\bibnamefont
  {Kato}},\ }\href {\doibase 10.1143/JPSJ.5.435} {\bibfield  {journal}
  {\bibinfo  {journal} {J. Phys. Soc. Jpn.}\ }\textbf {\bibinfo {volume} {5}},\
  \bibinfo {pages} {435} (\bibinfo {year} {1950})}\BibitemShut {NoStop}%
\bibitem [{\citenamefont {Messiah}(1961)}]{Messiah}%
  \BibitemOpen
  \bibfield  {author} {\bibinfo {author} {\bibfnamefont {A.}~\bibnamefont
  {Messiah}},\ }\href@noop {} {\emph {\bibinfo {title} {Quantum Mechanics}}}\
  (\bibinfo  {publisher} {North-Holland},\ \bibinfo {address} {Amsterdam},\
  \bibinfo {year} {1961})\BibitemShut {NoStop}%
\bibitem [{\citenamefont {Papageorgiou}\ and\ \citenamefont
  {Traub}(2013)}]{Papageorgiou_etal:2013}%
  \BibitemOpen
  \bibfield  {author} {\bibinfo {author} {\bibfnamefont {A.}~\bibnamefont
  {Papageorgiou}}\ and\ \bibinfo {author} {\bibfnamefont {J.~F.}\ \bibnamefont
  {Traub}},\ }\href {\doibase 10.1103/PhysRevA.88.022316} {\bibfield  {journal}
  {\bibinfo  {journal} {Phys. Rev. A}\ }\textbf {\bibinfo {volume} {88}},\
  \bibinfo {pages} {022316} (\bibinfo {year} {2013})}\BibitemShut {NoStop}%
\bibitem [{\citenamefont {R{\o}nnow}\ \emph {et~al.}(2014)\citenamefont
  {R{\o}nnow}, \citenamefont {Wang}, \citenamefont {Job}, \citenamefont
  {Boixo}, \citenamefont {Isakov}, \citenamefont {Wecker}, \citenamefont
  {Martinis}, \citenamefont {Lidar},\ and\ \citenamefont
  {Troyer}}]{Ronnow_etal:2014}%
  \BibitemOpen
  \bibfield  {author} {\bibinfo {author} {\bibfnamefont {T.~F.}\ \bibnamefont
  {R{\o}nnow}}, \bibinfo {author} {\bibfnamefont {Z.}~\bibnamefont {Wang}},
  \bibinfo {author} {\bibfnamefont {J.}~\bibnamefont {Job}}, \bibinfo {author}
  {\bibfnamefont {S.}~\bibnamefont {Boixo}}, \bibinfo {author} {\bibfnamefont
  {S.~V.}\ \bibnamefont {Isakov}}, \bibinfo {author} {\bibfnamefont
  {D.}~\bibnamefont {Wecker}}, \bibinfo {author} {\bibfnamefont {J.~M.}\
  \bibnamefont {Martinis}}, \bibinfo {author} {\bibfnamefont {D.~A.}\
  \bibnamefont {Lidar}}, \ and\ \bibinfo {author} {\bibfnamefont
  {M.}~\bibnamefont {Troyer}},\ }\href {\doibase 10.1126/science.1252319}
  {\bibfield  {journal} {\bibinfo  {journal} {Science}\ }\textbf {\bibinfo
  {volume} {345}},\ \bibinfo {pages} {420} (\bibinfo {year}
  {2014})}\BibitemShut {NoStop}%
\bibitem [{\citenamefont {Roland}\ and\ \citenamefont
  {Cerf}(2002)}]{Roland_etal:2002}%
  \BibitemOpen
  \bibfield  {author} {\bibinfo {author} {\bibfnamefont {J.}~\bibnamefont
  {Roland}}\ and\ \bibinfo {author} {\bibfnamefont {N.~J.}\ \bibnamefont
  {Cerf}},\ }\href {\doibase 10.1103/PhysRevA.65.042308} {\bibfield  {journal}
  {\bibinfo  {journal} {Phys. Rev. A}\ }\textbf {\bibinfo {volume} {65}},\
  \bibinfo {pages} {042308} (\bibinfo {year} {2002})}\BibitemShut {NoStop}%
\bibitem [{\citenamefont {Grover}(1997)}]{Grover:1997}%
  \BibitemOpen
  \bibfield  {author} {\bibinfo {author} {\bibfnamefont {L.~K.}\ \bibnamefont
  {Grover}},\ }\href {\doibase 10.1103/PhysRevLett.79.325} {\bibfield
  {journal} {\bibinfo  {journal} {Phys. Rev. Lett.}\ }\textbf {\bibinfo
  {volume} {79}},\ \bibinfo {pages} {325} (\bibinfo {year} {1997})}\BibitemShut
  {NoStop}%
\bibitem [{\citenamefont {Somma}\ \emph {et~al.}(2012)\citenamefont {Somma},
  \citenamefont {Nagaj},\ and\ \citenamefont {Kieferov\'a}}]{Somma_etal:2012}%
  \BibitemOpen
  \bibfield  {author} {\bibinfo {author} {\bibfnamefont {R.~D.}\ \bibnamefont
  {Somma}}, \bibinfo {author} {\bibfnamefont {D.}~\bibnamefont {Nagaj}}, \ and\
  \bibinfo {author} {\bibfnamefont {M.}~\bibnamefont {Kieferov\'a}},\ }\href
  {\doibase 10.1103/PhysRevLett.109.050501} {\bibfield  {journal} {\bibinfo
  {journal} {Phys. Rev. Lett.}\ }\textbf {\bibinfo {volume} {109}},\ \bibinfo
  {pages} {050501} (\bibinfo {year} {2012})}\BibitemShut {NoStop}%
\bibitem [{\citenamefont {Hen}(2014{\natexlab{a}})}]{Hen:2014a}%
  \BibitemOpen
  \bibfield  {author} {\bibinfo {author} {\bibfnamefont {I.}~\bibnamefont
  {Hen}},\ }\href {http://stacks.iop.org/1751-8121/47/i=4/a=045305} {\bibfield
  {journal} {\bibinfo  {journal} {J. Phys. A}\ }\textbf {\bibinfo {volume}
  {47}},\ \bibinfo {pages} {045305} (\bibinfo {year}
  {2014}{\natexlab{a}})}\BibitemShut {NoStop}%
\bibitem [{\citenamefont {Hen}(2014{\natexlab{b}})}]{Hen:2014b}%
  \BibitemOpen
  \bibfield  {author} {\bibinfo {author} {\bibfnamefont {I.}~\bibnamefont
  {Hen}},\ }\href {http://stacks.iop.org/0295-5075/105/i=5/a=50005} {\bibfield
  {journal} {\bibinfo  {journal} {Europhys. Lett.}\ }\textbf {\bibinfo {volume}
  {105}},\ \bibinfo {pages} {50005} (\bibinfo {year}
  {2014}{\natexlab{b}})}\BibitemShut {NoStop}%
\bibitem [{\citenamefont {Hen}(2016)}]{Hen:2016}%
  \BibitemOpen
  \bibfield  {author} {\bibinfo {author} {\bibfnamefont {I.}~\bibnamefont
  {Hen}},\ }\href {https://arxiv.org/abs/1612.06012} {\enquote {\bibinfo
  {title} {Realizable quantum adiabatic search},}\ } (\bibinfo {year} {2016}),\
  \Eprint {http://arxiv.org/abs/1612.06012} {arXiv:1612.06012} \BibitemShut
  {NoStop}%
\bibitem [{\citenamefont {Muthukrishnan}\ \emph {et~al.}(2016)\citenamefont
  {Muthukrishnan}, \citenamefont {Albash},\ and\ \citenamefont
  {Lidar}}]{Muthukrishnan_etal:2016}%
  \BibitemOpen
  \bibfield  {author} {\bibinfo {author} {\bibfnamefont {S.}~\bibnamefont
  {Muthukrishnan}}, \bibinfo {author} {\bibfnamefont {T.}~\bibnamefont
  {Albash}}, \ and\ \bibinfo {author} {\bibfnamefont {D.~A.}\ \bibnamefont
  {Lidar}},\ }\href {\doibase 10.1103/PhysRevX.6.031010} {\bibfield  {journal}
  {\bibinfo  {journal} {Phys. Rev. X}\ }\textbf {\bibinfo {volume} {6}},\
  \bibinfo {pages} {031010} (\bibinfo {year} {2016})}\BibitemShut {NoStop}%
\bibitem [{\citenamefont {Johnson}\ \emph {et~al.}(2011)\citenamefont
  {Johnson}, \citenamefont {Amin}, \citenamefont {Gildert}, \citenamefont
  {Lanting}, \citenamefont {Hamze}, \citenamefont {Dickson}, \citenamefont
  {Harris}, \citenamefont {Berkley}, \citenamefont {Johansson}, \citenamefont
  {Bunyk}, \citenamefont {Chapple}, \citenamefont {Enderud}, \citenamefont
  {Hilton}, \citenamefont {Karimi}, \citenamefont {Ladizinsky}, \citenamefont
  {Ladizinsky}, \citenamefont {Oh}, \citenamefont {Perminov}, \citenamefont
  {Rich}, \citenamefont {Thom}, \citenamefont {Tolkacheva}, \citenamefont
  {Truncik}, \citenamefont {Uchaikin}, \citenamefont {Wang}, \citenamefont
  {Wilson},\ and\ \citenamefont {Rose}}]{Johnson_etal:2011}%
  \BibitemOpen
  \bibfield  {author} {\bibinfo {author} {\bibfnamefont {M.~W.}\ \bibnamefont
  {Johnson}}, \bibinfo {author} {\bibfnamefont {M.~H.~S.}\ \bibnamefont
  {Amin}}, \bibinfo {author} {\bibfnamefont {S.}~\bibnamefont {Gildert}},
  \bibinfo {author} {\bibfnamefont {T.}~\bibnamefont {Lanting}}, \bibinfo
  {author} {\bibfnamefont {F.}~\bibnamefont {Hamze}}, \bibinfo {author}
  {\bibfnamefont {N.}~\bibnamefont {Dickson}}, \bibinfo {author} {\bibfnamefont
  {R.}~\bibnamefont {Harris}}, \bibinfo {author} {\bibfnamefont {A.~J.}\
  \bibnamefont {Berkley}}, \bibinfo {author} {\bibfnamefont {J.}~\bibnamefont
  {Johansson}}, \bibinfo {author} {\bibfnamefont {P.}~\bibnamefont {Bunyk}},
  \bibinfo {author} {\bibfnamefont {E.~M.}\ \bibnamefont {Chapple}}, \bibinfo
  {author} {\bibfnamefont {C.}~\bibnamefont {Enderud}}, \bibinfo {author}
  {\bibfnamefont {J.~P.}\ \bibnamefont {Hilton}}, \bibinfo {author}
  {\bibfnamefont {K.}~\bibnamefont {Karimi}}, \bibinfo {author} {\bibfnamefont
  {E.}~\bibnamefont {Ladizinsky}}, \bibinfo {author} {\bibfnamefont
  {N.}~\bibnamefont {Ladizinsky}}, \bibinfo {author} {\bibfnamefont
  {T.}~\bibnamefont {Oh}}, \bibinfo {author} {\bibfnamefont {I.}~\bibnamefont
  {Perminov}}, \bibinfo {author} {\bibfnamefont {C.}~\bibnamefont {Rich}},
  \bibinfo {author} {\bibfnamefont {M.~C.}\ \bibnamefont {Thom}}, \bibinfo
  {author} {\bibfnamefont {E.}~\bibnamefont {Tolkacheva}}, \bibinfo {author}
  {\bibfnamefont {C.~J.~S.}\ \bibnamefont {Truncik}}, \bibinfo {author}
  {\bibfnamefont {S.}~\bibnamefont {Uchaikin}}, \bibinfo {author}
  {\bibfnamefont {J.}~\bibnamefont {Wang}}, \bibinfo {author} {\bibfnamefont
  {B.}~\bibnamefont {Wilson}}, \ and\ \bibinfo {author} {\bibfnamefont
  {G.}~\bibnamefont {Rose}},\ }\href {http://dx.doi.org/10.1038/nature10012}
  {\bibfield  {journal} {\bibinfo  {journal} {Nature}\ }\textbf {\bibinfo
  {volume} {473}},\ \bibinfo {pages} {194} (\bibinfo {year}
  {2011})}\BibitemShut {NoStop}%
\bibitem [{\citenamefont {Lanting}\ \emph {et~al.}(2014)\citenamefont
  {Lanting}, \citenamefont {Przybysz}, \citenamefont {Smirnov}, \citenamefont
  {Spedalieri}, \citenamefont {Amin}, \citenamefont {Berkley}, \citenamefont
  {Harris}, \citenamefont {Altomare}, \citenamefont {Boixo}, \citenamefont
  {Bunyk}, \citenamefont {Dickson}, \citenamefont {Enderud}, \citenamefont
  {Hilton}, \citenamefont {Hoskinson}, \citenamefont {Johnson}, \citenamefont
  {Ladizinsky}, \citenamefont {Ladizinsky}, \citenamefont {Neufeld},
  \citenamefont {Oh}, \citenamefont {Perminov}, \citenamefont {Rich},
  \citenamefont {Thom}, \citenamefont {Tolkacheva}, \citenamefont {Uchaikin},
  \citenamefont {Wilson},\ and\ \citenamefont {Rose}}]{Lanting_etal:2014}%
  \BibitemOpen
  \bibfield  {author} {\bibinfo {author} {\bibfnamefont {T.}~\bibnamefont
  {Lanting}}, \bibinfo {author} {\bibfnamefont {A.~J.}\ \bibnamefont
  {Przybysz}}, \bibinfo {author} {\bibfnamefont {A.~Y.}\ \bibnamefont
  {Smirnov}}, \bibinfo {author} {\bibfnamefont {F.~M.}\ \bibnamefont
  {Spedalieri}}, \bibinfo {author} {\bibfnamefont {M.~H.}\ \bibnamefont
  {Amin}}, \bibinfo {author} {\bibfnamefont {A.~J.}\ \bibnamefont {Berkley}},
  \bibinfo {author} {\bibfnamefont {R.}~\bibnamefont {Harris}}, \bibinfo
  {author} {\bibfnamefont {F.}~\bibnamefont {Altomare}}, \bibinfo {author}
  {\bibfnamefont {S.}~\bibnamefont {Boixo}}, \bibinfo {author} {\bibfnamefont
  {P.}~\bibnamefont {Bunyk}}, \bibinfo {author} {\bibfnamefont
  {N.}~\bibnamefont {Dickson}}, \bibinfo {author} {\bibfnamefont
  {C.}~\bibnamefont {Enderud}}, \bibinfo {author} {\bibfnamefont {J.~P.}\
  \bibnamefont {Hilton}}, \bibinfo {author} {\bibfnamefont {E.}~\bibnamefont
  {Hoskinson}}, \bibinfo {author} {\bibfnamefont {M.~W.}\ \bibnamefont
  {Johnson}}, \bibinfo {author} {\bibfnamefont {E.}~\bibnamefont {Ladizinsky}},
  \bibinfo {author} {\bibfnamefont {N.}~\bibnamefont {Ladizinsky}}, \bibinfo
  {author} {\bibfnamefont {R.}~\bibnamefont {Neufeld}}, \bibinfo {author}
  {\bibfnamefont {T.}~\bibnamefont {Oh}}, \bibinfo {author} {\bibfnamefont
  {I.}~\bibnamefont {Perminov}}, \bibinfo {author} {\bibfnamefont
  {C.}~\bibnamefont {Rich}}, \bibinfo {author} {\bibfnamefont {M.~C.}\
  \bibnamefont {Thom}}, \bibinfo {author} {\bibfnamefont {E.}~\bibnamefont
  {Tolkacheva}}, \bibinfo {author} {\bibfnamefont {S.}~\bibnamefont
  {Uchaikin}}, \bibinfo {author} {\bibfnamefont {A.~B.}\ \bibnamefont
  {Wilson}}, \ and\ \bibinfo {author} {\bibfnamefont {G.}~\bibnamefont
  {Rose}},\ }\href {\doibase 10.1103/PhysRevX.4.021041} {\bibfield  {journal}
  {\bibinfo  {journal} {Phys. Rev. X}\ }\textbf {\bibinfo {volume} {4}},\
  \bibinfo {pages} {021041} (\bibinfo {year} {2014})}\BibitemShut {NoStop}%
\bibitem [{\citenamefont {Boixo}\ \emph {et~al.}(2014)\citenamefont {Boixo},
  \citenamefont {Ronnow}, \citenamefont {Isakov}, \citenamefont {Wang},
  \citenamefont {Wecker}, \citenamefont {Lidar}, \citenamefont {Martinis},\
  and\ \citenamefont {Troyer}}]{Boixo_etal:2014}%
  \BibitemOpen
  \bibfield  {author} {\bibinfo {author} {\bibfnamefont {S.}~\bibnamefont
  {Boixo}}, \bibinfo {author} {\bibfnamefont {T.~F.}\ \bibnamefont {Ronnow}},
  \bibinfo {author} {\bibfnamefont {S.~V.}\ \bibnamefont {Isakov}}, \bibinfo
  {author} {\bibfnamefont {Z.}~\bibnamefont {Wang}}, \bibinfo {author}
  {\bibfnamefont {D.}~\bibnamefont {Wecker}}, \bibinfo {author} {\bibfnamefont
  {D.~A.}\ \bibnamefont {Lidar}}, \bibinfo {author} {\bibfnamefont {J.~M.}\
  \bibnamefont {Martinis}}, \ and\ \bibinfo {author} {\bibfnamefont
  {M.}~\bibnamefont {Troyer}},\ }\href {\doibase 10.1038/nphys2900} {\bibfield
  {journal} {\bibinfo  {journal} {Nat. Phys.}\ }\textbf {\bibinfo {volume}
  {10}},\ \bibinfo {pages} {218} (\bibinfo {year} {2014})}\BibitemShut
  {NoStop}%
\bibitem [{\citenamefont {Katzgraber}\ \emph {et~al.}(2015)\citenamefont
  {Katzgraber}, \citenamefont {Hamze}, \citenamefont {Zhu}, \citenamefont
  {Ochoa},\ and\ \citenamefont {Munoz-Bauza}}]{Katzgraber:2015}%
  \BibitemOpen
  \bibfield  {author} {\bibinfo {author} {\bibfnamefont {H.~G.}\ \bibnamefont
  {Katzgraber}}, \bibinfo {author} {\bibfnamefont {F.}~\bibnamefont {Hamze}},
  \bibinfo {author} {\bibfnamefont {Z.}~\bibnamefont {Zhu}}, \bibinfo {author}
  {\bibfnamefont {A.~J.}\ \bibnamefont {Ochoa}}, \ and\ \bibinfo {author}
  {\bibfnamefont {H.}~\bibnamefont {Munoz-Bauza}},\ }\href {\doibase
  10.1103/PhysRevX.5.031026} {\bibfield  {journal} {\bibinfo  {journal} {Phys.
  Rev. X}\ }\textbf {\bibinfo {volume} {5}},\ \bibinfo {pages} {031026}
  (\bibinfo {year} {2015})}\BibitemShut {NoStop}%
\bibitem [{\citenamefont {Denchev}\ \emph {et~al.}(2016)\citenamefont
  {Denchev}, \citenamefont {Boixo}, \citenamefont {Isakov}, \citenamefont
  {Ding}, \citenamefont {Babbush}, \citenamefont {Smelyanskiy}, \citenamefont
  {Martinis},\ and\ \citenamefont {Neven}}]{Denchev_etal:2016}%
  \BibitemOpen
  \bibfield  {author} {\bibinfo {author} {\bibfnamefont {V.~S.}\ \bibnamefont
  {Denchev}}, \bibinfo {author} {\bibfnamefont {S.}~\bibnamefont {Boixo}},
  \bibinfo {author} {\bibfnamefont {S.~V.}\ \bibnamefont {Isakov}}, \bibinfo
  {author} {\bibfnamefont {N.}~\bibnamefont {Ding}}, \bibinfo {author}
  {\bibfnamefont {R.}~\bibnamefont {Babbush}}, \bibinfo {author} {\bibfnamefont
  {V.}~\bibnamefont {Smelyanskiy}}, \bibinfo {author} {\bibfnamefont
  {J.}~\bibnamefont {Martinis}}, \ and\ \bibinfo {author} {\bibfnamefont
  {H.}~\bibnamefont {Neven}},\ }\href {\doibase 10.1103/PhysRevX.6.031015}
  {\bibfield  {journal} {\bibinfo  {journal} {Phys. Rev. X}\ }\textbf {\bibinfo
  {volume} {6}},\ \bibinfo {pages} {031015} (\bibinfo {year}
  {2016})}\BibitemShut {NoStop}%
\bibitem [{\citenamefont {Altshuler}\ \emph {et~al.}(2010)\citenamefont
  {Altshuler}, \citenamefont {Krovi},\ and\ \citenamefont
  {Roland}}]{Altshuler_etal:2010}%
  \BibitemOpen
  \bibfield  {author} {\bibinfo {author} {\bibfnamefont {B.}~\bibnamefont
  {Altshuler}}, \bibinfo {author} {\bibfnamefont {H.}~\bibnamefont {Krovi}}, \
  and\ \bibinfo {author} {\bibfnamefont {J.}~\bibnamefont {Roland}},\ }\href
  {\doibase 10.1073/pnas.1002116107} {\bibfield  {journal} {\bibinfo  {journal}
  {Proc. Natl. Acad. Sci. U.S.A.}\ }\textbf {\bibinfo {volume} {107}},\
  \bibinfo {pages} {12446} (\bibinfo {year} {2010})}\BibitemShut {NoStop}%
\bibitem [{\citenamefont {Farhi}\ \emph {et~al.}(2011)\citenamefont {Farhi},
  \citenamefont {Goldston}, \citenamefont {Gosset}, \citenamefont {Gutmann},
  \citenamefont {Meyer},\ and\ \citenamefont {Shor}}]{Farhi_etal:2011}%
  \BibitemOpen
  \bibfield  {author} {\bibinfo {author} {\bibfnamefont {E.}~\bibnamefont
  {Farhi}}, \bibinfo {author} {\bibfnamefont {J.}~\bibnamefont {Goldston}},
  \bibinfo {author} {\bibfnamefont {D.}~\bibnamefont {Gosset}}, \bibinfo
  {author} {\bibfnamefont {S.}~\bibnamefont {Gutmann}}, \bibinfo {author}
  {\bibfnamefont {H.~B.}\ \bibnamefont {Meyer}}, \ and\ \bibinfo {author}
  {\bibfnamefont {P.}~\bibnamefont {Shor}},\ }\href
  {http://dl.acm.org/citation.cfm?id=2011395.2011396} {\bibfield  {journal}
  {\bibinfo  {journal} {Quant. Info. Comput.}\ }\textbf {\bibinfo {volume}
  {11}},\ \bibinfo {pages} {181} (\bibinfo {year} {2011})}\BibitemShut
  {NoStop}%
\bibitem [{\citenamefont {Choi}(2010)}]{Choi:2010}%
  \BibitemOpen
  \bibfield  {author} {\bibinfo {author} {\bibfnamefont {V.}~\bibnamefont
  {Choi}},\ }\href {https://arxiv.org/abs/1004.2226} {\enquote {\bibinfo
  {title} {Adiabatic quantum algorithms for the np-complete maximum-weight
  independent set, exact cover and 3sat problems},}\ } (\bibinfo {year}
  {2010}),\ \Eprint {http://arxiv.org/abs/1004.2226} {arXiv:1004.2226}
  \BibitemShut {NoStop}%
\bibitem [{\citenamefont {Dickson}\ and\ \citenamefont
  {Amin}(2011)}]{Dickson_etal:2011}%
  \BibitemOpen
  \bibfield  {author} {\bibinfo {author} {\bibfnamefont {N.~G.}\ \bibnamefont
  {Dickson}}\ and\ \bibinfo {author} {\bibfnamefont {M.~H.~S.}\ \bibnamefont
  {Amin}},\ }\href {\doibase 10.1103/PhysRevLett.106.050502} {\bibfield
  {journal} {\bibinfo  {journal} {Phys. Rev. Lett.}\ }\textbf {\bibinfo
  {volume} {106}},\ \bibinfo {pages} {050502} (\bibinfo {year}
  {2011})}\BibitemShut {NoStop}%
\bibitem [{\citenamefont {Dickson}(2011)}]{Dickson:2011}%
  \BibitemOpen
  \bibfield  {author} {\bibinfo {author} {\bibfnamefont {N.~G.}\ \bibnamefont
  {Dickson}},\ }\href {http://stacks.iop.org/1367-2630/13/i=7/a=073011}
  {\bibfield  {journal} {\bibinfo  {journal} {New J. Phys.}\ }\textbf {\bibinfo
  {volume} {13}},\ \bibinfo {pages} {073011} (\bibinfo {year}
  {2011})}\BibitemShut {NoStop}%
\bibitem [{\citenamefont {Jozsa}\ and\ \citenamefont
  {Linden}(2003)}]{Jozsa_etal:2003}%
  \BibitemOpen
  \bibfield  {author} {\bibinfo {author} {\bibfnamefont {R.}~\bibnamefont
  {Jozsa}}\ and\ \bibinfo {author} {\bibfnamefont {N.}~\bibnamefont {Linden}},\
  }\href {\doibase 10.1098/rspa.2002.1097} {\bibfield  {journal} {\bibinfo
  {journal} {Proc. R. Soc. A}\ }\textbf {\bibinfo {volume} {459}},\ \bibinfo
  {pages} {2011} (\bibinfo {year} {2003})}\BibitemShut {NoStop}%
\bibitem [{\citenamefont {Vidal}(2003)}]{Vidal:2003}%
  \BibitemOpen
  \bibfield  {author} {\bibinfo {author} {\bibfnamefont {G.}~\bibnamefont
  {Vidal}},\ }\href {\doibase 10.1103/PhysRevLett.91.147902} {\bibfield
  {journal} {\bibinfo  {journal} {Phys. Rev. Lett.}\ }\textbf {\bibinfo
  {volume} {91}},\ \bibinfo {pages} {147902} (\bibinfo {year}
  {2003})}\BibitemShut {NoStop}%
\bibitem [{\citenamefont {Parker}\ and\ \citenamefont
  {Plenio}(2002)}]{Parker_etal:2002}%
  \BibitemOpen
  \bibfield  {author} {\bibinfo {author} {\bibfnamefont {S.}~\bibnamefont
  {Parker}}\ and\ \bibinfo {author} {\bibfnamefont {M.~B.}\ \bibnamefont
  {Plenio}},\ }\href {\doibase 10.1080/09500340110107207} {\bibfield  {journal}
  {\bibinfo  {journal} {Journal of Modern Optics}\ }\textbf {\bibinfo {volume}
  {49}},\ \bibinfo {pages} {1325} (\bibinfo {year} {2002})}\BibitemShut
  {NoStop}%
\bibitem [{\citenamefont {Shimoni}\ \emph {et~al.}(2004)\citenamefont
  {Shimoni}, \citenamefont {Shapira},\ and\ \citenamefont
  {Biham}}]{Shimoni_etal:2004}%
  \BibitemOpen
  \bibfield  {author} {\bibinfo {author} {\bibfnamefont {Y.}~\bibnamefont
  {Shimoni}}, \bibinfo {author} {\bibfnamefont {D.}~\bibnamefont {Shapira}}, \
  and\ \bibinfo {author} {\bibfnamefont {O.}~\bibnamefont {Biham}},\ }\href
  {\doibase 10.1103/PhysRevA.69.062303} {\bibfield  {journal} {\bibinfo
  {journal} {Phys. Rev. A}\ }\textbf {\bibinfo {volume} {69}},\ \bibinfo
  {pages} {062303} (\bibinfo {year} {2004})}\BibitemShut {NoStop}%
\bibitem [{\citenamefont {Shimoni}\ \emph {et~al.}(2005)\citenamefont
  {Shimoni}, \citenamefont {Shapira},\ and\ \citenamefont
  {Biham}}]{Shimoni_etal:2005}%
  \BibitemOpen
  \bibfield  {author} {\bibinfo {author} {\bibfnamefont {Y.}~\bibnamefont
  {Shimoni}}, \bibinfo {author} {\bibfnamefont {D.}~\bibnamefont {Shapira}}, \
  and\ \bibinfo {author} {\bibfnamefont {O.}~\bibnamefont {Biham}},\ }\href
  {\doibase 10.1103/PhysRevA.72.062308} {\bibfield  {journal} {\bibinfo
  {journal} {Phys. Rev. A}\ }\textbf {\bibinfo {volume} {72}},\ \bibinfo
  {pages} {062308} (\bibinfo {year} {2005})}\BibitemShut {NoStop}%
\bibitem [{\citenamefont {Or\'us}\ and\ \citenamefont
  {Latorre}(2004)}]{Orus_etal:2004}%
  \BibitemOpen
  \bibfield  {author} {\bibinfo {author} {\bibfnamefont {R.}~\bibnamefont
  {Or\'us}}\ and\ \bibinfo {author} {\bibfnamefont {J.~I.}\ \bibnamefont
  {Latorre}},\ }\href {\doibase 10.1103/PhysRevA.69.052308} {\bibfield
  {journal} {\bibinfo  {journal} {Phys. Rev. A}\ }\textbf {\bibinfo {volume}
  {69}},\ \bibinfo {pages} {052308} (\bibinfo {year} {2004})}\BibitemShut
  {NoStop}%
\bibitem [{\citenamefont {Ukena}\ and\ \citenamefont
  {Shimizu}(2004)}]{Ukena_etal:2004}%
  \BibitemOpen
  \bibfield  {author} {\bibinfo {author} {\bibfnamefont {A.}~\bibnamefont
  {Ukena}}\ and\ \bibinfo {author} {\bibfnamefont {A.}~\bibnamefont
  {Shimizu}},\ }\href {\doibase 10.1103/PhysRevA.69.022301} {\bibfield
  {journal} {\bibinfo  {journal} {Phys. Rev. A}\ }\textbf {\bibinfo {volume}
  {69}},\ \bibinfo {pages} {022301} (\bibinfo {year} {2004})}\BibitemShut
  {NoStop}%
\bibitem [{\citenamefont {Shimizu}\ \emph {et~al.}(2013)\citenamefont
  {Shimizu}, \citenamefont {Matsuzaki},\ and\ \citenamefont
  {Ukena}}]{Shimizu_etal:2013}%
  \BibitemOpen
  \bibfield  {author} {\bibinfo {author} {\bibfnamefont {A.}~\bibnamefont
  {Shimizu}}, \bibinfo {author} {\bibfnamefont {Y.}~\bibnamefont {Matsuzaki}},
  \ and\ \bibinfo {author} {\bibfnamefont {A.}~\bibnamefont {Ukena}},\ }\href
  {\doibase 10.7566/JPSJ.82.054801} {\bibfield  {journal} {\bibinfo  {journal}
  {J. Phys. Soc. Jpn.}\ }\textbf {\bibinfo {volume} {82}},\ \bibinfo {pages}
  {054801} (\bibinfo {year} {2013})}\BibitemShut {NoStop}%
\bibitem [{\citenamefont {Latorre}\ and\ \citenamefont
  {Or\'us}(2004)}]{Latorre_etal:2004}%
  \BibitemOpen
  \bibfield  {author} {\bibinfo {author} {\bibfnamefont {J.~I.}\ \bibnamefont
  {Latorre}}\ and\ \bibinfo {author} {\bibfnamefont {R.}~\bibnamefont
  {Or\'us}},\ }\href {\doibase 10.1103/PhysRevA.69.062302} {\bibfield
  {journal} {\bibinfo  {journal} {Phys. Rev. A}\ }\textbf {\bibinfo {volume}
  {69}},\ \bibinfo {pages} {062302} (\bibinfo {year} {2004})}\BibitemShut
  {NoStop}%
\bibitem [{\citenamefont {Rezakhani}\ \emph {et~al.}(2009)\citenamefont
  {Rezakhani}, \citenamefont {Kuo}, \citenamefont {Hamma}, \citenamefont
  {Lidar},\ and\ \citenamefont {Zanardi}}]{Rezakhani_etal:2009}%
  \BibitemOpen
  \bibfield  {author} {\bibinfo {author} {\bibfnamefont {A.~T.}\ \bibnamefont
  {Rezakhani}}, \bibinfo {author} {\bibfnamefont {W.-J.}\ \bibnamefont {Kuo}},
  \bibinfo {author} {\bibfnamefont {A.}~\bibnamefont {Hamma}}, \bibinfo
  {author} {\bibfnamefont {D.~A.}\ \bibnamefont {Lidar}}, \ and\ \bibinfo
  {author} {\bibfnamefont {P.}~\bibnamefont {Zanardi}},\ }\href {\doibase
  10.1103/PhysRevLett.103.080502} {\bibfield  {journal} {\bibinfo  {journal}
  {Phys. Rev. Lett.}\ }\textbf {\bibinfo {volume} {103}},\ \bibinfo {pages}
  {080502} (\bibinfo {year} {2009})}\BibitemShut {NoStop}%
\bibitem [{\citenamefont {Bauer}\ \emph {et~al.}(2015)\citenamefont {Bauer},
  \citenamefont {Wang}, \citenamefont {Pizorn},\ and\ \citenamefont
  {Troyer}}]{Bauer_etal:2015}%
  \BibitemOpen
  \bibfield  {author} {\bibinfo {author} {\bibfnamefont {B.}~\bibnamefont
  {Bauer}}, \bibinfo {author} {\bibfnamefont {L.}~\bibnamefont {Wang}},
  \bibinfo {author} {\bibfnamefont {I.}~\bibnamefont {Pizorn}}, \ and\ \bibinfo
  {author} {\bibfnamefont {M.}~\bibnamefont {Troyer}},\ }\href
  {https://arxiv.org/abs/1501.06914} {\enquote {\bibinfo {title} {Entanglement
  as a resource in adiabatic quantum optimization},}\ } (\bibinfo {year}
  {2015}),\ \Eprint {http://arxiv.org/abs/1501.06914} {arXiv:1501.06914}
  \BibitemShut {NoStop}%
\bibitem [{\citenamefont {Hauke}\ \emph {et~al.}(2015)\citenamefont {Hauke},
  \citenamefont {Bonnes}, \citenamefont {Heyl},\ and\ \citenamefont
  {Lechner}}]{Hauke_etal:2015}%
  \BibitemOpen
  \bibfield  {author} {\bibinfo {author} {\bibfnamefont {P.}~\bibnamefont
  {Hauke}}, \bibinfo {author} {\bibfnamefont {L.}~\bibnamefont {Bonnes}},
  \bibinfo {author} {\bibfnamefont {M.}~\bibnamefont {Heyl}}, \ and\ \bibinfo
  {author} {\bibfnamefont {W.}~\bibnamefont {Lechner}},\ }\href {\doibase
  10.3389/fphy.2015.00021} {\bibfield  {journal} {\bibinfo  {journal}
  {Frontiers in Physics}\ }\textbf {\bibinfo {volume} {3}},\ \bibinfo {pages}
  {21} (\bibinfo {year} {2015})}\BibitemShut {NoStop}%
\bibitem [{\citenamefont {Batle}\ \emph {et~al.}(2016)\citenamefont {Batle},
  \citenamefont {Ooi}, \citenamefont {Farouk}, \citenamefont {Abutalib},\ and\
  \citenamefont {Abdalla}}]{Batle_etal:2016}%
  \BibitemOpen
  \bibfield  {author} {\bibinfo {author} {\bibfnamefont {J.}~\bibnamefont
  {Batle}}, \bibinfo {author} {\bibfnamefont {C.~H.~R.}\ \bibnamefont {Ooi}},
  \bibinfo {author} {\bibfnamefont {A.}~\bibnamefont {Farouk}}, \bibinfo
  {author} {\bibfnamefont {M.}~\bibnamefont {Abutalib}}, \ and\ \bibinfo
  {author} {\bibfnamefont {S.}~\bibnamefont {Abdalla}},\ }\href {\doibase
  10.1007/s11128-016-1324-x} {\bibfield  {journal} {\bibinfo  {journal} {Quant.
  Info. Proc.}\ }\textbf {\bibinfo {volume} {15}},\ \bibinfo {pages} {3081}
  (\bibinfo {year} {2016})}\BibitemShut {NoStop}%
\bibitem [{\citenamefont {Albash}\ \emph {et~al.}(2015)\citenamefont {Albash},
  \citenamefont {Hen}, \citenamefont {Spedalieri},\ and\ \citenamefont
  {Lidar}}]{Albash_etal:2015}%
  \BibitemOpen
  \bibfield  {author} {\bibinfo {author} {\bibfnamefont {T.}~\bibnamefont
  {Albash}}, \bibinfo {author} {\bibfnamefont {I.}~\bibnamefont {Hen}},
  \bibinfo {author} {\bibfnamefont {F.~M.}\ \bibnamefont {Spedalieri}}, \ and\
  \bibinfo {author} {\bibfnamefont {D.~A.}\ \bibnamefont {Lidar}},\ }\href
  {\doibase 10.1103/PhysRevA.92.062328} {\bibfield  {journal} {\bibinfo
  {journal} {Phys. Rev. A}\ }\textbf {\bibinfo {volume} {92}},\ \bibinfo
  {pages} {062328} (\bibinfo {year} {2015})}\BibitemShut {NoStop}%
\bibitem [{\citenamefont {Wei}\ and\ \citenamefont
  {Ying}(2006)}]{Wei_etal:2006}%
  \BibitemOpen
  \bibfield  {author} {\bibinfo {author} {\bibfnamefont {Z.}~\bibnamefont
  {Wei}}\ and\ \bibinfo {author} {\bibfnamefont {M.}~\bibnamefont {Ying}},\
  }\href {\doibase https://doi.org/10.1016/j.physleta.2006.01.098} {\bibfield
  {journal} {\bibinfo  {journal} {Phys. Lett. A}\ }\textbf {\bibinfo {volume}
  {354}},\ \bibinfo {pages} {271 } (\bibinfo {year} {2006})}\BibitemShut
  {NoStop}%
\bibitem [{\citenamefont {Knuth}(1976)}]{Knuth:1976}%
  \BibitemOpen
  \bibfield  {author} {\bibinfo {author} {\bibfnamefont {D.~E.}\ \bibnamefont
  {Knuth}},\ }\href {\doibase 10.1145/1008328.1008329} {\bibfield  {journal}
  {\bibinfo  {journal} {SIGACT News}\ }\textbf {\bibinfo {volume} {8}},\
  \bibinfo {pages} {18} (\bibinfo {year} {1976})}\BibitemShut {NoStop}%
\bibitem [{\citenamefont {Cormen}\ \emph {et~al.}(2009)\citenamefont {Cormen},
  \citenamefont {Leiserson}, \citenamefont {Rivest},\ and\ \citenamefont
  {Stein}}]{Cormen_etal}%
  \BibitemOpen
  \bibfield  {author} {\bibinfo {author} {\bibfnamefont {T.~H.}\ \bibnamefont
  {Cormen}}, \bibinfo {author} {\bibfnamefont {C.~E.}\ \bibnamefont
  {Leiserson}}, \bibinfo {author} {\bibfnamefont {R.~L.}\ \bibnamefont
  {Rivest}}, \ and\ \bibinfo {author} {\bibfnamefont {C.}~\bibnamefont
  {Stein}},\ }\href@noop {} {\emph {\bibinfo {title} {Introduction to
  Algorithms}}},\ \bibinfo {edition} {3rd}\ ed.\ (\bibinfo  {publisher} {MIT
  Press},\ \bibinfo {address} {Cambridge MA},\ \bibinfo {year}
  {2009})\BibitemShut {NoStop}%
\bibitem [{\citenamefont {Nielsen}\ and\ \citenamefont
  {Chuang}(2000)}]{NielsenChuang}%
  \BibitemOpen
  \bibfield  {author} {\bibinfo {author} {\bibfnamefont {M.~A.}\ \bibnamefont
  {Nielsen}}\ and\ \bibinfo {author} {\bibfnamefont {I.~L.}\ \bibnamefont
  {Chuang}},\ }\href@noop {} {\emph {\bibinfo {title} {Quantum Computation and
  Quantum Information}}}\ (\bibinfo  {publisher} {Cambridge University Press},\
  \bibinfo {address} {Cambridge},\ \bibinfo {year} {2000})\BibitemShut
  {NoStop}%
\bibitem [{\citenamefont {Morimae}\ \emph {et~al.}(2005)\citenamefont
  {Morimae}, \citenamefont {Sugita},\ and\ \citenamefont
  {Shimizu}}]{Morimae_etal:2005}%
  \BibitemOpen
  \bibfield  {author} {\bibinfo {author} {\bibfnamefont {T.}~\bibnamefont
  {Morimae}}, \bibinfo {author} {\bibfnamefont {A.}~\bibnamefont {Sugita}}, \
  and\ \bibinfo {author} {\bibfnamefont {A.}~\bibnamefont {Shimizu}},\ }\href
  {\doibase 10.1103/PhysRevA.71.032317} {\bibfield  {journal} {\bibinfo
  {journal} {Phys. Rev. A}\ }\textbf {\bibinfo {volume} {71}},\ \bibinfo
  {pages} {032317} (\bibinfo {year} {2005})}\BibitemShut {NoStop}%
\bibitem [{\citenamefont {Sugita}\ and\ \citenamefont
  {Shimizu}(2005)}]{Sugita_etal:2005}%
  \BibitemOpen
  \bibfield  {author} {\bibinfo {author} {\bibfnamefont {A.}~\bibnamefont
  {Sugita}}\ and\ \bibinfo {author} {\bibfnamefont {A.}~\bibnamefont
  {Shimizu}},\ }\href {\doibase 10.1143/JPSJ.74.1883} {\bibfield  {journal}
  {\bibinfo  {journal} {J. Phys. Soc. Jpn}\ }\textbf {\bibinfo {volume} {74}},\
  \bibinfo {pages} {1883} (\bibinfo {year} {2005})}\BibitemShut {NoStop}%
\bibitem [{\citenamefont {Fr\"{o}wis}\ and\ \citenamefont
  {D\"{u}r}(2012)}]{Frowis_etal:2012}%
  \BibitemOpen
  \bibfield  {author} {\bibinfo {author} {\bibfnamefont {F.}~\bibnamefont
  {Fr\"{o}wis}}\ and\ \bibinfo {author} {\bibfnamefont {W.}~\bibnamefont
  {D\"{u}r}},\ }\href {http://stacks.iop.org/1367-2630/14/i=9/a=093039}
  {\bibfield  {journal} {\bibinfo  {journal} {New J. Phys.}\ }\textbf {\bibinfo
  {volume} {14}},\ \bibinfo {pages} {093039} (\bibinfo {year}
  {2012})}\BibitemShut {NoStop}%
\bibitem [{\citenamefont {Yadin}\ and\ \citenamefont
  {Vedral}(2015)}]{Yadin:2015}%
  \BibitemOpen
  \bibfield  {author} {\bibinfo {author} {\bibfnamefont {B.}~\bibnamefont
  {Yadin}}\ and\ \bibinfo {author} {\bibfnamefont {V.}~\bibnamefont {Vedral}},\
  }\href {\doibase 10.1103/PhysRevA.92.022356} {\bibfield  {journal} {\bibinfo
  {journal} {Phys. Rev. A}\ }\textbf {\bibinfo {volume} {92}},\ \bibinfo
  {pages} {022356} (\bibinfo {year} {2015})}\BibitemShut {NoStop}%
\bibitem [{\citenamefont {Tichy}\ \emph {et~al.}(2016)\citenamefont {Tichy},
  \citenamefont {Park}, \citenamefont {Kang}, \citenamefont {Jeong},\ and\
  \citenamefont {M\o{}lmer}}]{Tichy_etal:2016}%
  \BibitemOpen
  \bibfield  {author} {\bibinfo {author} {\bibfnamefont {M.~C.}\ \bibnamefont
  {Tichy}}, \bibinfo {author} {\bibfnamefont {C.-Y.}\ \bibnamefont {Park}},
  \bibinfo {author} {\bibfnamefont {M.}~\bibnamefont {Kang}}, \bibinfo {author}
  {\bibfnamefont {H.}~\bibnamefont {Jeong}}, \ and\ \bibinfo {author}
  {\bibfnamefont {K.}~\bibnamefont {M\o{}lmer}},\ }\href {\doibase
  10.1103/PhysRevA.93.042314} {\bibfield  {journal} {\bibinfo  {journal} {Phys.
  Rev. A}\ }\textbf {\bibinfo {volume} {93}},\ \bibinfo {pages} {042314}
  (\bibinfo {year} {2016})}\BibitemShut {NoStop}%
\bibitem [{\citenamefont {Abad}\ and\ \citenamefont
  {Karimipour}(2016)}]{Abad_etal:2016}%
  \BibitemOpen
  \bibfield  {author} {\bibinfo {author} {\bibfnamefont {T.}~\bibnamefont
  {Abad}}\ and\ \bibinfo {author} {\bibfnamefont {V.}~\bibnamefont
  {Karimipour}},\ }\href {\doibase 10.1103/PhysRevB.93.195127} {\bibfield
  {journal} {\bibinfo  {journal} {Phys. Rev. B}\ }\textbf {\bibinfo {volume}
  {93}},\ \bibinfo {pages} {195127} (\bibinfo {year} {2016})}\BibitemShut
  {NoStop}%
\bibitem [{\citenamefont {Park}\ \emph {et~al.}(2016)\citenamefont {Park},
  \citenamefont {Kang}, \citenamefont {Lee}, \citenamefont {Bang},
  \citenamefont {Lee},\ and\ \citenamefont {Jeong}}]{Park_etal:2016}%
  \BibitemOpen
  \bibfield  {author} {\bibinfo {author} {\bibfnamefont {C.-Y.}\ \bibnamefont
  {Park}}, \bibinfo {author} {\bibfnamefont {M.}~\bibnamefont {Kang}}, \bibinfo
  {author} {\bibfnamefont {C.-W.}\ \bibnamefont {Lee}}, \bibinfo {author}
  {\bibfnamefont {J.}~\bibnamefont {Bang}}, \bibinfo {author} {\bibfnamefont
  {S.-W.}\ \bibnamefont {Lee}}, \ and\ \bibinfo {author} {\bibfnamefont
  {H.}~\bibnamefont {Jeong}},\ }\href {\doibase 10.1103/PhysRevA.94.052105}
  {\bibfield  {journal} {\bibinfo  {journal} {Phys. Rev. A}\ }\textbf {\bibinfo
  {volume} {94}},\ \bibinfo {pages} {052105} (\bibinfo {year}
  {2016})}\BibitemShut {NoStop}%
\bibitem [{\citenamefont {Kuwahara}\ \emph {et~al.}(2017)\citenamefont
  {Kuwahara}, \citenamefont {Arad}, \citenamefont {Amico},\ and\ \citenamefont
  {Vedral}}]{Kuwahara_etal:2017}%
  \BibitemOpen
  \bibfield  {author} {\bibinfo {author} {\bibfnamefont {T.}~\bibnamefont
  {Kuwahara}}, \bibinfo {author} {\bibfnamefont {I.}~\bibnamefont {Arad}},
  \bibinfo {author} {\bibfnamefont {L.}~\bibnamefont {Amico}}, \ and\ \bibinfo
  {author} {\bibfnamefont {V.}~\bibnamefont {Vedral}},\ }\href
  {http://stacks.iop.org/2058-9565/2/i=1/a=015005} {\bibfield  {journal}
  {\bibinfo  {journal} {Quantum Science and Technology}\ }\textbf {\bibinfo
  {volume} {2}},\ \bibinfo {pages} {015005} (\bibinfo {year}
  {2017})}\BibitemShut {NoStop}%
\bibitem [{\citenamefont {Tatsuta}\ and\ \citenamefont
  {Shimizu}(2017)}]{Tatsuta_etal:2017}%
  \BibitemOpen
  \bibfield  {author} {\bibinfo {author} {\bibfnamefont {M.}~\bibnamefont
  {Tatsuta}}\ and\ \bibinfo {author} {\bibfnamefont {A.}~\bibnamefont
  {Shimizu}},\ }\href {https://arxiv.org/abs/1703.05034} {\enquote {\bibinfo
  {title} {Conversion of thermal equilibrium states into superpositions of
  macroscopically distinct states},}\ } (\bibinfo {year} {2017}),\ \Eprint
  {http://arxiv.org/abs/1703.05034} {arXiv:1703.05034} \BibitemShut {NoStop}%
\bibitem [{\citenamefont {Morimae}\ and\ \citenamefont
  {Shimizu}(2006)}]{Morimae_etal:2006}%
  \BibitemOpen
  \bibfield  {author} {\bibinfo {author} {\bibfnamefont {T.}~\bibnamefont
  {Morimae}}\ and\ \bibinfo {author} {\bibfnamefont {A.}~\bibnamefont
  {Shimizu}},\ }\href {\doibase 10.1103/PhysRevA.74.052111} {\bibfield
  {journal} {\bibinfo  {journal} {Phys. Rev. A}\ }\textbf {\bibinfo {volume}
  {74}},\ \bibinfo {pages} {052111} (\bibinfo {year} {2006})}\BibitemShut
  {NoStop}%
\bibitem [{\citenamefont {Aharonov}\ \emph {et~al.}(2007)\citenamefont
  {Aharonov}, \citenamefont {van Dam}, \citenamefont {Kempe}, \citenamefont
  {Landau}, \citenamefont {Lloyd},\ and\ \citenamefont
  {Regev}}]{Aharonov_etal:2007}%
  \BibitemOpen
  \bibfield  {author} {\bibinfo {author} {\bibfnamefont {D.}~\bibnamefont
  {Aharonov}}, \bibinfo {author} {\bibfnamefont {W.}~\bibnamefont {van Dam}},
  \bibinfo {author} {\bibfnamefont {J.}~\bibnamefont {Kempe}}, \bibinfo
  {author} {\bibfnamefont {Z.}~\bibnamefont {Landau}}, \bibinfo {author}
  {\bibfnamefont {S.}~\bibnamefont {Lloyd}}, \ and\ \bibinfo {author}
  {\bibfnamefont {O.}~\bibnamefont {Regev}},\ }\href {\doibase
  10.1137/S0097539705447323} {\bibfield  {journal} {\bibinfo  {journal} {SIAM
  J. Comput.}\ }\textbf {\bibinfo {volume} {37}},\ \bibinfo {pages} {166}
  (\bibinfo {year} {2007})}\BibitemShut {NoStop}%
\bibitem [{\citenamefont {Kempe}\ \emph {et~al.}(2006)\citenamefont {Kempe},
  \citenamefont {Kitaev},\ and\ \citenamefont {Regev}}]{Kempe_etal:2006}%
  \BibitemOpen
  \bibfield  {author} {\bibinfo {author} {\bibfnamefont {J.}~\bibnamefont
  {Kempe}}, \bibinfo {author} {\bibfnamefont {A.}~\bibnamefont {Kitaev}}, \
  and\ \bibinfo {author} {\bibfnamefont {O.}~\bibnamefont {Regev}},\ }\href
  {\doibase 10.1137/S0097539704445226} {\bibfield  {journal} {\bibinfo
  {journal} {SIAM J. Comput.}\ }\textbf {\bibinfo {volume} {35}},\ \bibinfo
  {pages} {1070} (\bibinfo {year} {2006})}\BibitemShut {NoStop}%
\bibitem [{\citenamefont {Mizel}\ \emph {et~al.}(2007)\citenamefont {Mizel},
  \citenamefont {Lidar},\ and\ \citenamefont {Mitchell}}]{Mizel_etal:2007}%
  \BibitemOpen
  \bibfield  {author} {\bibinfo {author} {\bibfnamefont {A.}~\bibnamefont
  {Mizel}}, \bibinfo {author} {\bibfnamefont {D.~A.}\ \bibnamefont {Lidar}}, \
  and\ \bibinfo {author} {\bibfnamefont {M.}~\bibnamefont {Mitchell}},\ }\href
  {\doibase 10.1103/PhysRevLett.99.070502} {\bibfield  {journal} {\bibinfo
  {journal} {Phys. Rev. Lett.}\ }\textbf {\bibinfo {volume} {99}},\ \bibinfo
  {pages} {070502} (\bibinfo {year} {2007})}\BibitemShut {NoStop}%
\bibitem [{\citenamefont {Gosset}\ \emph {et~al.}(2015)\citenamefont {Gosset},
  \citenamefont {Terhal},\ and\ \citenamefont {Vershynina}}]{Gosset_etal:2015}%
  \BibitemOpen
  \bibfield  {author} {\bibinfo {author} {\bibfnamefont {D.}~\bibnamefont
  {Gosset}}, \bibinfo {author} {\bibfnamefont {B.~M.}\ \bibnamefont {Terhal}},
  \ and\ \bibinfo {author} {\bibfnamefont {A.}~\bibnamefont {Vershynina}},\
  }\href {\doibase 10.1103/PhysRevLett.114.140501} {\bibfield  {journal}
  {\bibinfo  {journal} {Phys. Rev. Lett.}\ }\textbf {\bibinfo {volume} {114}},\
  \bibinfo {pages} {140501} (\bibinfo {year} {2015})}\BibitemShut {NoStop}%
\bibitem [{\citenamefont {Van~den Nest}(2013)}]{Nest:2013}%
  \BibitemOpen
  \bibfield  {author} {\bibinfo {author} {\bibfnamefont {M.}~\bibnamefont
  {Van~den Nest}},\ }\href {\doibase 10.1103/PhysRevLett.110.060504} {\bibfield
   {journal} {\bibinfo  {journal} {Phys. Rev. Lett.}\ }\textbf {\bibinfo
  {volume} {110}},\ \bibinfo {pages} {060504} (\bibinfo {year}
  {2013})}\BibitemShut {NoStop}%
\bibitem [{\citenamefont {Bennett}\ \emph {et~al.}(1997)\citenamefont
  {Bennett}, \citenamefont {Bernstein}, \citenamefont {Brassard},\ and\
  \citenamefont {Vazirani}}]{Bennett_etal:1997}%
  \BibitemOpen
  \bibfield  {author} {\bibinfo {author} {\bibfnamefont {C.~H.}\ \bibnamefont
  {Bennett}}, \bibinfo {author} {\bibfnamefont {E.}~\bibnamefont {Bernstein}},
  \bibinfo {author} {\bibfnamefont {G.}~\bibnamefont {Brassard}}, \ and\
  \bibinfo {author} {\bibfnamefont {U.}~\bibnamefont {Vazirani}},\ }\href
  {\doibase 10.1137/S0097539796300933} {\bibfield  {journal} {\bibinfo
  {journal} {SIAM J. Comput.}\ }\textbf {\bibinfo {volume} {26}},\ \bibinfo
  {pages} {1510} (\bibinfo {year} {1997})}\BibitemShut {NoStop}%
\bibitem [{\citenamefont {Mermin}(2007)}]{Mermin}%
  \BibitemOpen
  \bibfield  {author} {\bibinfo {author} {\bibfnamefont {N.~D.}\ \bibnamefont
  {Mermin}},\ }\href@noop {} {\emph {\bibinfo {title} {Quantum Computer
  Science: An Introduction}}}\ (\bibinfo  {publisher} {Cambridge University
  Press},\ \bibinfo {address} {Cambridge},\ \bibinfo {year} {2007})\BibitemShut
  {NoStop}%
\bibitem [{\citenamefont {Jansen}\ \emph {et~al.}(2007)\citenamefont {Jansen},
  \citenamefont {Ruskai},\ and\ \citenamefont {Seiler}}]{Jansen_etal:2007}%
  \BibitemOpen
  \bibfield  {author} {\bibinfo {author} {\bibfnamefont {S.}~\bibnamefont
  {Jansen}}, \bibinfo {author} {\bibfnamefont {M.-B.}\ \bibnamefont {Ruskai}},
  \ and\ \bibinfo {author} {\bibfnamefont {R.}~\bibnamefont {Seiler}},\ }\href
  {\doibase 10.1063/1.2798382} {\bibfield  {journal} {\bibinfo  {journal} {J.
  Math. Phys.}\ }\textbf {\bibinfo {volume} {48}},\ \bibinfo {pages} {102111}
  (\bibinfo {year} {2007})}\BibitemShut {NoStop}%
\bibitem [{\citenamefont {Deutsch}\ and\ \citenamefont
  {Jozsa}(1992)}]{Deutsch_etal:1992}%
  \BibitemOpen
  \bibfield  {author} {\bibinfo {author} {\bibfnamefont {D.}~\bibnamefont
  {Deutsch}}\ and\ \bibinfo {author} {\bibfnamefont {R.}~\bibnamefont
  {Jozsa}},\ }\href {\doibase 10.1098/rspa.1992.0167} {\bibfield  {journal}
  {\bibinfo  {journal} {Proc. R. Soc. A}\ }\textbf {\bibinfo {volume} {439}},\
  \bibinfo {pages} {553} (\bibinfo {year} {1992})}\BibitemShut {NoStop}%
\bibitem [{\citenamefont {Cleve}\ \emph {et~al.}(1998)\citenamefont {Cleve},
  \citenamefont {Ekert}, \citenamefont {Macchiavello},\ and\ \citenamefont
  {Mosca}}]{Cleve_etal:1998}%
  \BibitemOpen
  \bibfield  {author} {\bibinfo {author} {\bibfnamefont {R.}~\bibnamefont
  {Cleve}}, \bibinfo {author} {\bibfnamefont {A.}~\bibnamefont {Ekert}},
  \bibinfo {author} {\bibfnamefont {C.}~\bibnamefont {Macchiavello}}, \ and\
  \bibinfo {author} {\bibfnamefont {M.}~\bibnamefont {Mosca}},\ }\href
  {\doibase 10.1098/rspa.1998.0164} {\bibfield  {journal} {\bibinfo  {journal}
  {Proc. R. Soc. A}\ }\textbf {\bibinfo {volume} {454}},\ \bibinfo {pages}
  {339} (\bibinfo {year} {1998})}\BibitemShut {NoStop}%
\bibitem [{\citenamefont {Sarandy}\ and\ \citenamefont
  {Lidar}(2005)}]{Sarandy_etal:2005}%
  \BibitemOpen
  \bibfield  {author} {\bibinfo {author} {\bibfnamefont {M.~S.}\ \bibnamefont
  {Sarandy}}\ and\ \bibinfo {author} {\bibfnamefont {D.~A.}\ \bibnamefont
  {Lidar}},\ }\href {\doibase 10.1103/PhysRevLett.95.250503} {\bibfield
  {journal} {\bibinfo  {journal} {Phys. Rev. Lett.}\ }\textbf {\bibinfo
  {volume} {95}},\ \bibinfo {pages} {250503} (\bibinfo {year}
  {2005})}\BibitemShut {NoStop}%
\bibitem [{\citenamefont {Bernstein}\ and\ \citenamefont
  {Vazirani}(1997)}]{Bernstein_etal:1997}%
  \BibitemOpen
  \bibfield  {author} {\bibinfo {author} {\bibfnamefont {E.}~\bibnamefont
  {Bernstein}}\ and\ \bibinfo {author} {\bibfnamefont {U.}~\bibnamefont
  {Vazirani}},\ }\href {\doibase 10.1137/S0097539796300921} {\bibfield
  {journal} {\bibinfo  {journal} {SIAM J. Comput.}\ }\textbf {\bibinfo {volume}
  {26}},\ \bibinfo {pages} {1411} (\bibinfo {year} {1997})}\BibitemShut
  {NoStop}%
\bibitem [{\citenamefont {Simon}(1997)}]{Simon:1997}%
  \BibitemOpen
  \bibfield  {author} {\bibinfo {author} {\bibfnamefont {D.~R.}\ \bibnamefont
  {Simon}},\ }\href {\doibase 10.1137/S0097539796298637} {\bibfield  {journal}
  {\bibinfo  {journal} {SIAM J. Comput.}\ }\textbf {\bibinfo {volume} {26}},\
  \bibinfo {pages} {1474} (\bibinfo {year} {1997})}\BibitemShut {NoStop}%
\bibitem [{\citenamefont {Childs}\ \emph {et~al.}(2003)\citenamefont {Childs},
  \citenamefont {Cleve}, \citenamefont {Deotto}, \citenamefont {Farhi},
  \citenamefont {Gutmann},\ and\ \citenamefont {Spielman}}]{Childs_etal:2003}%
  \BibitemOpen
  \bibfield  {author} {\bibinfo {author} {\bibfnamefont {A.~M.}\ \bibnamefont
  {Childs}}, \bibinfo {author} {\bibfnamefont {R.}~\bibnamefont {Cleve}},
  \bibinfo {author} {\bibfnamefont {E.}~\bibnamefont {Deotto}}, \bibinfo
  {author} {\bibfnamefont {E.}~\bibnamefont {Farhi}}, \bibinfo {author}
  {\bibfnamefont {S.}~\bibnamefont {Gutmann}}, \ and\ \bibinfo {author}
  {\bibfnamefont {D.~A.}\ \bibnamefont {Spielman}},\ }in\ \href {\doibase
  10.1145/780542.780552} {\emph {\bibinfo {booktitle} {Proceedings of the
  Thirty-fifth Annual ACM Symposium on Theory of Computing}}},\ \bibinfo
  {series and number} {STOC '03}\ (\bibinfo  {publisher} {ACM},\ \bibinfo
  {address} {New York, NY, USA},\ \bibinfo {year} {2003})\ pp.\ \bibinfo
  {pages} {59--68}\BibitemShut {NoStop}%
\bibitem [{\citenamefont {Young}\ \emph {et~al.}(2010)\citenamefont {Young},
  \citenamefont {Knysh},\ and\ \citenamefont {Smelyanskiy}}]{Young_etal:2010}%
  \BibitemOpen
  \bibfield  {author} {\bibinfo {author} {\bibfnamefont {A.~P.}\ \bibnamefont
  {Young}}, \bibinfo {author} {\bibfnamefont {S.}~\bibnamefont {Knysh}}, \ and\
  \bibinfo {author} {\bibfnamefont {V.~N.}\ \bibnamefont {Smelyanskiy}},\
  }\href {\doibase 10.1103/PhysRevLett.104.020502} {\bibfield  {journal}
  {\bibinfo  {journal} {Phys. Rev. Lett.}\ }\textbf {\bibinfo {volume} {104}},\
  \bibinfo {pages} {020502} (\bibinfo {year} {2010})}\BibitemShut {NoStop}%
\bibitem [{\citenamefont {Shimizu}\ and\ \citenamefont
  {Miyadera}(2002)}]{Shimizu_etal:2002}%
  \BibitemOpen
  \bibfield  {author} {\bibinfo {author} {\bibfnamefont {A.}~\bibnamefont
  {Shimizu}}\ and\ \bibinfo {author} {\bibfnamefont {T.}~\bibnamefont
  {Miyadera}},\ }\href {\doibase 10.1103/PhysRevLett.89.270403} {\bibfield
  {journal} {\bibinfo  {journal} {Phys. Rev. Lett.}\ }\textbf {\bibinfo
  {volume} {89}},\ \bibinfo {pages} {270403} (\bibinfo {year}
  {2002})}\BibitemShut {NoStop}%
\bibitem [{\citenamefont {Shimizu}\ and\ \citenamefont
  {Morimae}(2005)}]{Shimizu_etal:2005}%
  \BibitemOpen
  \bibfield  {author} {\bibinfo {author} {\bibfnamefont {A.}~\bibnamefont
  {Shimizu}}\ and\ \bibinfo {author} {\bibfnamefont {T.}~\bibnamefont
  {Morimae}},\ }\href {\doibase 10.1103/PhysRevLett.95.090401} {\bibfield
  {journal} {\bibinfo  {journal} {Phys. Rev. Lett.}\ }\textbf {\bibinfo
  {volume} {95}},\ \bibinfo {pages} {090401} (\bibinfo {year}
  {2005})}\BibitemShut {NoStop}%
\bibitem [{\citenamefont {Amin}\ \emph {et~al.}(2008)\citenamefont {Amin},
  \citenamefont {Love},\ and\ \citenamefont {Truncik}}]{Amin_etal:2008}%
  \BibitemOpen
  \bibfield  {author} {\bibinfo {author} {\bibfnamefont {M.~H.~S.}\
  \bibnamefont {Amin}}, \bibinfo {author} {\bibfnamefont {P.~J.}\ \bibnamefont
  {Love}}, \ and\ \bibinfo {author} {\bibfnamefont {C.~J.~S.}\ \bibnamefont
  {Truncik}},\ }\href {\doibase 10.1103/PhysRevLett.100.060503} {\bibfield
  {journal} {\bibinfo  {journal} {Phys. Rev. Lett.}\ }\textbf {\bibinfo
  {volume} {100}},\ \bibinfo {pages} {060503} (\bibinfo {year}
  {2008})}\BibitemShut {NoStop}%
\bibitem [{\citenamefont {Wang}\ and\ \citenamefont {Guo}(1989)}]{WangGuo}%
  \BibitemOpen
  \bibfield  {author} {\bibinfo {author} {\bibfnamefont {Z.~X.}\ \bibnamefont
  {Wang}}\ and\ \bibinfo {author} {\bibfnamefont {D.~R.}\ \bibnamefont {Guo}},\
  }\href@noop {} {\emph {\bibinfo {title} {Special Functions}}}\ (\bibinfo
  {publisher} {World Scientific},\ \bibinfo {address} {Singapore},\ \bibinfo
  {year} {1989})\BibitemShut {NoStop}%
\end{thebibliography}
\end{document}